# MLGaze: Machine Learning-Based Analysis of Gaze Error Patterns in Consumer Eye Tracking Systems

**Anuradha Kar**

École Normale Supérieure de Lyon, 46 Allée d'Italie, 69007 Lyon, France; anuradha.kar@ens-lyon.fr or anuradha.kar49@gmail.com; Tel.: +33-6238-20074

**Abstract:** Analyzing the gaze accuracy characteristics of an eye tracker is a critical task as its gaze data is frequently affected by non-ideal operating conditions in various consumer eye tracking applications. In previous research on pattern analysis of gaze data, efforts were made to model human visual behaviors and cognitive processes. What remains relatively unexplored are questions related to identifying gaze error sources as well as quantifying and modeling their impacts on the data quality of eye trackers. In this study, gaze error patterns produced by a commercial eye tracking device were studied with the help of machine learning algorithms, such as classifiers and regression models. Gaze data were collected from a group of participants under multiple conditions that commonly affect eye trackers operating on desktop and handheld platforms. These conditions (referred here as error sources) include user distance, head pose, and eye-tracker pose variations, and the collected gaze data were used to train the classifier and regression models. It was seen that while the impact of the different error sources on gaze data characteristics were nearly impossible to distinguish by visual inspection or from data statistics, machine learning models were successful in identifying the impact of the different error sources and predicting the variability in gaze error levels due to these conditions. The objective of this study was to investigate the efficacy of machine learning methods towards the detection and prediction of gaze error patterns, which would enable an in-depth understanding of the data quality and reliability of eye trackers under unconstrained operating conditions. Coding resources for all the machine learning methods adopted in this study were included in an open repository named MLGaze to allow researchers to replicate the principles presented here using data from their own eye trackers.

## 1. Introduction

*1.1. Research Questions and Motivation*

Gaze data obtained from eye trackers operating on various consumer platforms is frequently affected by a multitude of factors (or error sources), such as the head pose, user distance, display properties of the setup, illumination variations, and occlusions. The impact of these factors on gaze data are manifested in the form of gaze estimation errors whose characteristics or distributions have not been explored adequately in contemporary gaze research [1]. Conventionally, researchers attempt to improve the accuracy of eye trackers or calibration methods, while gaze error patterns are rarely analyzed and thus there remain many questions regarding the nature of gaze estimation errors. For example, it is not known whether the above error sources produce any particular pattern of errors or if the nature of gaze errors follows any statistical distribution or if they are simply random. These aspects cannot be understood by looking at raw gaze data, which is almost always corrupted with noise and outliers, or even by studying mean gaze error values.

With respect to gaze error analysis, several questions arise: These include: (1) how can gaze errors caused by one error source be distinguished from those caused by another? (2) How can the presence of different error sources in a certain gaze dataset be detected without prior knowledge? (3) Is it possible to predict the level of gaze errors that have been caused by different error sources? (4) Can suitable features be extracted from gaze datasets to identify the different error sources. These questions form the main topics that are addressed in this paper.

This paper focuses on defining methods for detailed analysis of eye tracking data obtained from a generic commercial eye tracker for the detection, identification, and prediction of gaze errors. The aim of this study was to observe gaze error patterns produced by three error sources that commonly affect eye trackers in consumer platforms like desktop and tablets. These include head movement, user distance, and platform orientation. It was found that there currently exists no publicly available gaze dataset that contains raw gaze and ground truth (gaze target location) data along with signatures of these error sources. Therefore, a new eye tracking dataset was built by collecting gaze data from 20 participants using a commercial eye tracker on both a desktop and a tablet platform. This dataset was published in an open repository for use by the gaze research community (link to the dataset webpage is in Section 2.5.5). During data collection, variations in head pose, user distance, and platform orientations were introduced sequentially and in a calibrated manner so that the gaze data contains the influence of one known condition (or error source) at a time. Reference data, which does not have the influence of any of these conditions, was collected as well.

For the detection of gaze error patterns caused by the different error sources, the collected data was used for training machine learning (ML) algorithms by creating training features from the datasets. As the operating conditions change, the feature variables in the training data get affected and as a result the ML models learn to differentiate between error patterns caused by different sources. In this way, anomalous gaze data may be distinguished using classifiers, which were trained using both affected and reference gaze datasets. Finally, regression algorithms were built to model and predict gaze errors that may be produced by the above three error sources.

*1.2. Background and Scope*

Pattern classification and modelling approaches have been applied on eye tracking data for various purposes. In [2], the effect of the video frame rate on the viewing behavior and visual perception of users was studied by comparing gaze data patterns collected for high and low video frame rates. In [3], a new hybrid fuzzy approach was introduced to distinguish gaze patterns when users performed face and text scanning. In [4], the dominant gaze characteristics of experienced and inexperienced train drivers were classified using the Markov cluster (MCL) algorithm, and marked differences were observed between gaze patterns of the two driver classes. The authors in [5] used a Bayesian mixture model to learn the gaze behavior of drivers who perform a variety of tasks during conditionally automated driving, by classifying the driver's fixations and saccades. Gaze patterns were used with clustering and classification algorithms to predict the user's intention to perform a set of tasks in [6].

A special de-noising and segmentation algorithm based on naive segmented linear regression and hidden Markov models was developed in [7] to classify gaze features into fixations, smooth pursuits, saccades, and post-saccadic oscillations using good-quality as well as noisy data. Automated classification of fixations and saccades was achieved using random forest classifiers on high-quality as well as noisy gaze datasets in [8].

It may be observed that gaze or eye movement pattern modelling have mostly been applied towards either cognitive studies, i.e., to interpret viewing patterns or distinguish between oculomotor event types, e.g., saccades, fixations, and smooth pursuits. With respect to studying gaze error patterns, however, only a few works were reported. Examples include the work in [9], which estimated the two-dimensional gaze error distribution and used a predictive model to shift the gaze according to a directional error estimate. This is based on a previous study [10], which proposed a gaze estimation error model that can predict gaze error by taking into consideration factors, such as the mapping of pupil positions to scene camera coordinates, marker-based display detection, and mapping from a scene camera to on-screen gaze coordinates. The major difference between the approaches taken in [9,10] and the current study is that the authors in those papers modelled the gaze error based on the contributions from the components of the gaze estimation process, e.g., mapping of the pupil to the scene camera coordinates and that from the scene camera to the on-screen coordinates. However, in this study, the gaze errors were studied as impacts of operating conditions that are external to the gaze estimation system. The reason for adopting this approach in this study was that for most consumer eye trackers, the gaze estimation principle and components of the eye trackers are not accessible for inspection to users; however, the influence of the operating conditions is manifested via variations in the gaze data characteristics. Therefore, it is feasible to model the gaze errors of eye trackers by studying their output data characteristics in response to different operating conditions. It was also found that no research works till now have covered the aspect of the classification of gaze error patterns induced by different operating conditions and prediction of error levels under their influence. These form the basis of the concepts that are presented in this paper.

The identification of unusual or anomalous data patterns has motivated research in areas, such as computer networks, business intelligence, and data mining [11]. This study was inspired by research in these fields and applied machine learning principles, which have been used successfully in anomaly detection algorithms. Automatic detection of anomalous data has manifold advantages, for example, in identifying the presence of defects, and getting insights into system behavior while also helping in reducing the number of system failures and improving the chances of recovery [12]. Similar to the above research areas, eye tracking devices also generate bulk amounts of data and the data quality is critical for the use of the collected datasets. It is therefore imperative for researchers to ensure that over the periods of experiments, the gaze data follows consistent accuracy patterns. However, several factors, as described in this paper, arising from the user or experimental setup conditions can result in altered data characteristics, which must be detected and corrected to maintain the consistency of gaze research results. This forms the motivation behind the gaze error pattern identification and modelling work presented in this paper.

In this study, the term gaze error is frequently used, which is the angular difference between the gaze locations estimated by an eye tracker and the actual locations of the visual targets (also called ground truth), typically appearing on a display screen or viewing area. The "gaze error patterns" described here signify the magnitude distributions of the gaze error values over a given number of samples or display area. Error sources refer to non-ideal operating conditions, such as a high degree of head pose variation with respect to the frontal pose, too long or too short user distances, or conditions where an eye tracking platform is tilted to different angles as opposed to their neutral positions. Such conditions were considered in this study because they occur frequently during real world operations of eye trackers. However, it is not known to what extent these conditions affect the gaze data characteristics and impact the gaze error patterns.

The impact of these error sources on the gaze error patterns produced by a given eye tracker was studied in this paper. The scope of the paper was limited to studying one error source or operating condition at a time (i.e., either the head pose or user distance or platform pose was varied at a time). However, eye trackers could face more complex scenarios, for example, two or more error sources could occur together and affect an eye tracker's data. Such analysis was not covered in this study as firstly, relevant data was not available, and secondly, the occurrences of such complex circumstances are rarer than the ones considered here. It is likely that at a time there is only one error source affecting an eye tracker with a major influence compared to other sources, which may or may not be present. For example, the user distance is fixed in a desktop or automotive-based eye tracking systems and head motion is the major source of gaze tracking error in these platforms. The analysis in this paper was based on data collected from static remote eye tracker setups on desktop and tablet platforms.

*1.3. Organization of Contents*

The paper is organized as follows. Section 2 describes the gaze data collection setup and procedure for data pre-processing and exploration. Section 3 presents the gaze error detection using machine learning models. Section 4 presents the regression algorithms for the modelling of gaze errors. Section 5 describes the code repository for the implementation of the methods described in this paper.

**2. Experimental Methodology and Data Exploration**

The concepts of this study were implemented in several phases. Eye tracking data was collected through special experiments using two consumer platforms, a desktop and a tablet, under different operating conditions. The collected data went through the processing pipeline (Figure 1) with an initial investigation using statistical methods and visualizations before being fed to the machine learning models. The data collection experiments, experimental setup, and steps of the data analysis workflow are described below. All subjects gave their informed consent for inclusion before they participated in the study. The study was conducted in accordance with the Declaration of Helsinki, and the protocol was approved by the Ethics Committee of SFI Project ID: 13/SPP/I2868.

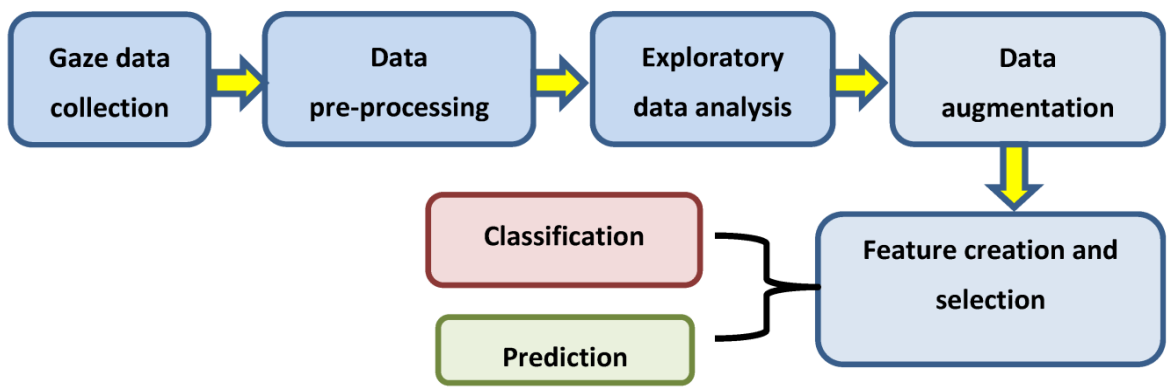

**Figure 1.** Gaze data processing and analysis pipeline followed in this study.

*2.1. Eye Tracking Data Collection*

A detailed description of the gaze data collection process and setup using a commercial eye tracker was provided in [1] and also in the sections below. Gaze data was collected from a group of 20 participants and for four user distances from both the desktop and tablet platforms.

*2.2. Eye Tracking UI and Device*

For eye tracking data collection, an eye tracker coupled with a visual stimulus interface (also called UI) was used. The trackers used were a Tobii EyeX 4C and an Eye tribe remote eye tracking device, which come with their own calibration routines. During an experiment session, participants were seated in front of the tracker with a chin rest and the UI ran simultaneously with the eye tracker. The UI shows a moving dot, which sequentially traces a grid of (5 × 3) known locations (also called the area of interest or AOI) over the desktop or tablet's display. The dot radius is 10 pixels and it stops at each AOI for 3 s before moving to the next, and gaze data was collected while participants looked at the dot. Gaze data comprised of a participant's gaze coordinates (x, y positions in pixels) on the (desktop/tablet) display and corresponding time stamps as estimated by the eye tracker. The known on-screen dot locations form the ground truth data and were used for accuracy calculations. The chin rest was used for stabilizing the participant's head for all experiments. A nine-point eye calibration was performed for all participants before each experiment session. Photos from the desktop and tablet experimental setup are shown in Figure 2a and Figure 2b, respectively.

*2.3. Setup and Experiment Details*

2.3.1. Setup Description

The desktop setup consisted of the eye tracker mounted on the screen of a desktop computer (Figure 2a). The screen diagonal was 22 inches (model: Asus VW22ATL–LCD), with a pixel resolution of 1680 × 1050. Two experiments were done with this setup. These were: (a) User distance experiments: In these, gaze data was collected at user-eye tracker distances of 50, 60, 70, and 80 cm; and (b) head-pose variability experiments: Head pose here refers to the position of a user's head in 3-D space in terms of the roll, pitch, and yaw (RPY) angles. During the experiments, a user was seated at a fixed distance (60 cm) from the tracker and was asked to vary their head position to distinct pose (RPY) angles each time, with respect to the frontal position (RPY = 0) while looking at the UI on the display. Their gaze was tracked on the UI and their head position was tracked simultaneously using an active appearance model [13], or AAM, which is a computer vision-based software that measures head pose RPY angles with 1 degree accuracy. To run the AAM, a Logitech HD Webcam C270 (1280 × 720px) was used to capture the facial video of users from which the head pose was estimated by the model. Sample images from the AAM showing various head pose angles are in Figure 2c. A sample video showing head pose estimation using AAM is in the supplementary materials with this paper. Neutral pose data is the same as the user distance data at 60 cm.

The tablet model was a Lenovo MIIX 310 with a screen diagonal size of 10.1 inches and pixel resolution of 1920 × 800. For the tablet experiments, the tablet was mounted on a gimbal tripod as shown in Figure 2d and the orientations were measured using the readings from the inbuilt inertial sensors of the tablet. For the tablet setup, two sets of experiments were dome with the same test UI as used for the desktop. The first were the user distance experiments done in the same way as for the desktop platform. The other was done by studying the impact of a variable platform orientation on the eye tracker data. For this, the orientation of the combined tablet-eye tracker setup was varied to known platform roll, pitch, and yaw angles (20 degree in each of roll, pitch yaw directions) as shown in Figure 2d. The user's head and distance from the tablet-tracker setup were kept fixed at 60cm. Gaze data was collected for each tablet orientation. Data for the neutral orientation was the same as gaze data at 60 cm from the tablet. The full list of experiments is shown in Figure 2e.

The maximum head and platform pose allowed by the eye tracker was 20 degrees and no tracker data could be obtained at distances below 50cm or above 80cm. Data from the user distance experiments were called UD50, UD60, UD70, and UD80 for distances of 50, 60, 70, and 80 cm, respectively. "HP" was used to denote head pose experiments. The process flow of experiments (Figure 2h) comprised of positioning the user in front of the desktop or tablet setup and calibrating their eyes with the tracker calibration software. Next, the UI was run with its data-logging routine to record gaze data in comma-separated values or CSV files, along with millisecond timestamps.

To position a user on the desktop and tablet setup, the center of the user's head was kept aligned with the screen center line as shown in Figure 2i. The height of the chin rest was adjusted to keep the user eyes level with the center of the display screen. The user was made to sit at increasing distances along the center line while keeping their head aligned with the center of the display.

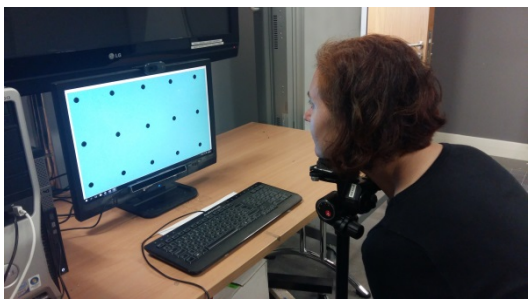 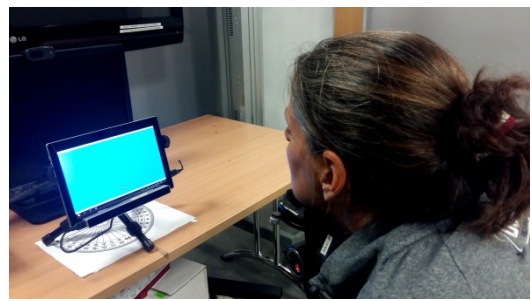

(**a**) (**b**)

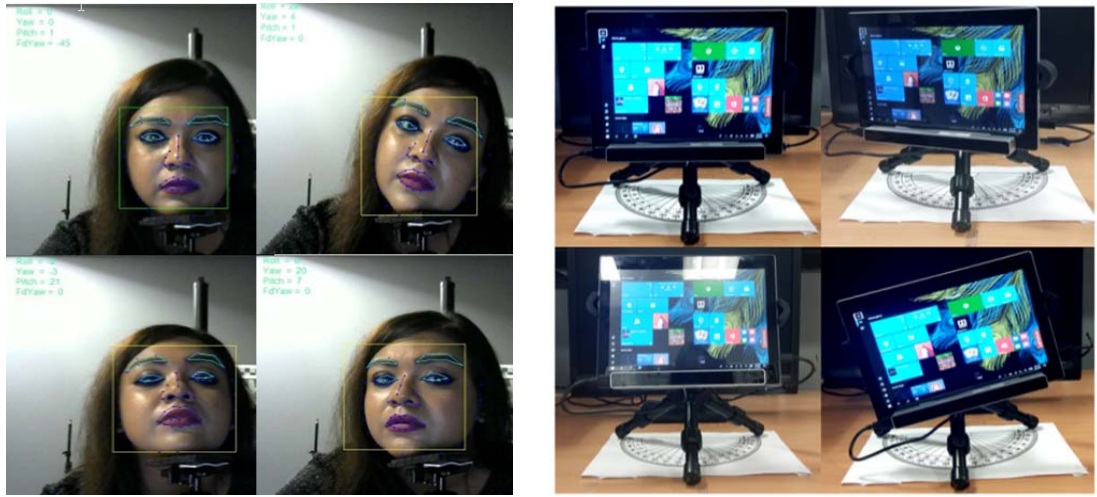

(c)                (d)

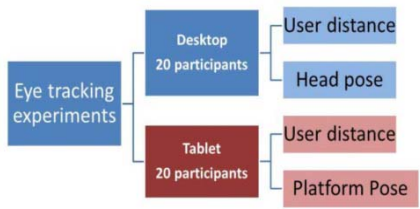     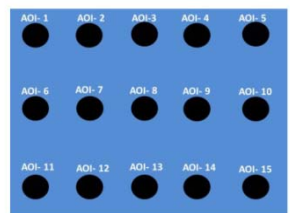

(e)                (f)

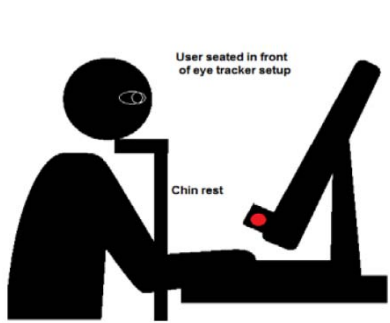     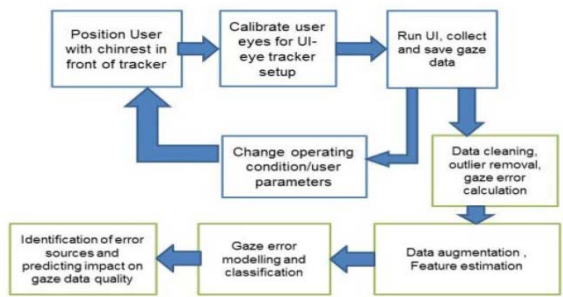

(g)                (h)

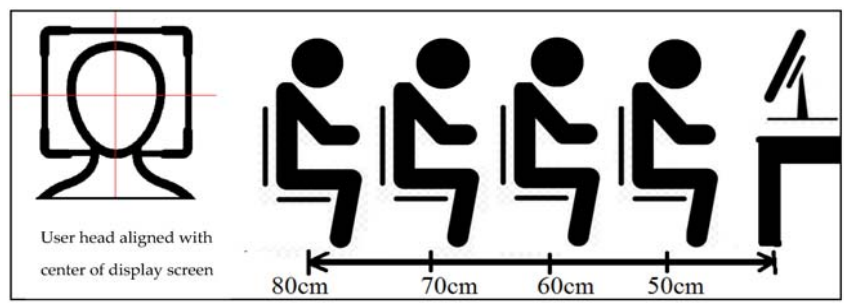

(i)

**Figure 2.** (**a**) Desktop setup and (**b**) tablet setup for gaze data collection using eye tracker. (**c**) Desktop setup and (**d**) variation of the tablet orientation for the experiments. (**e**) List of the experiments performed in this study. (**f**) Layout of the user interface (UI) for gaze data collection. (**g**) The user-eye tracker setup. (**h**) Process flow for data collection and analysis. (**i**) Schematic diagram showing the positioning of users during the experiments.

2.3.2. Participants and Demography Information

For both the desktop and tablet experiments, the same group of 20 participants (15 males, 5 females) were involved in the data collection for all experiments. The age range of all the users was between 30 and 45 years with a median age of 38 years. This study was done with participants not wearing glasses, in an indoor space and under uniform illumination levels, to rule out the impact of occlusion and illumination changes on the gaze data. The users were made familiar with the eye tracking setup through instructions and pilot experiments before the main data collection process was started.

*2.4. Data Preparation*

Figure 3 below shows the raw gaze data overlaid on the ground truth (or target locations), as obtained from the different experiments described above. It can be seen that it is impossible to decipher error patterns from gaze data by simply looking at it in raw form, and data from very different operating conditions often look similar and vice versa. However, as will be shown next, the eye tracking data obtained under different operating conditions do have diverse error distributions and statistical properties.

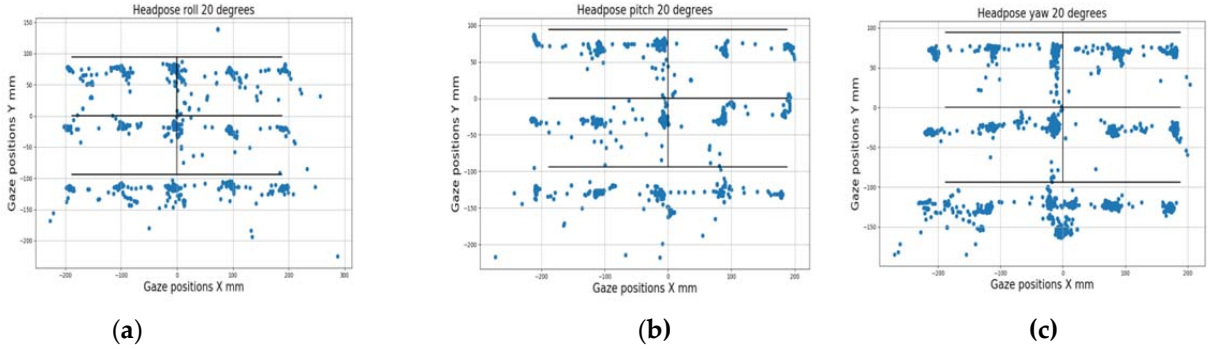

(**a**)         (**b**)         (**c**)

**Figure 3.** Raw gaze data (blue points) overlaid on ground truth data (black lines) for (a) head pose with roll of 20 degrees (**b**), head pitch of 20 degrees, and (**c**) head yaw 20 degrees. It can be seen that it is very hard to visually distinguish between data from different experiments done under different conditions, unless special pre-processing steps and learning algorithms are employed to identify the error source affecting the eye tracker.

The first step to prepare the gaze data for the learning tasks was to convert the raw gaze data coordinates (in pixels) into frontal gaze angles and gaze yaw and pitch angles. The ground truth data comprised of the screen locations (x, y coordinates in pixels) at which the black dot in the stimuli UI stopped during the gaze data collection (Figure 2f). The raw gaze x,y pixel coordinates of the left and right eye ($X_{left}$, $Y_{left}$, and $X_{right}$, $Y_{right}$ respectively) obtained from the tracker were used to estimate the gaze angle and gaze yaw and pitch angles as follows [1]:

$$GazeX = mean(\frac{x_{left} + x_{right}}{2}), \quad GazeY = mean(\frac{y_{left} + y_{right}}{2}). \tag{1}$$

The on-screen distance (OSD) of a user's gaze point was the distance between the origin and a certain gaze point with the coordinates *(GazeX, GazeY)*. In our case, the tracker was attached directly below the screens and the origin was the center of the screen. Therefore:

$$OSD(mm) = \mu\sqrt{(GazeX)^2 + (GazeY)^2}, \tag{2}$$

where $\mu$ is the pixel pitch of the display where the gaze was tracked in units of mm/pixel.

The gaze angle of a point on the screen relative to a user's eyes was calculated as:

$$Gaze\ angle(\theta gaze) = \tan^{-1}(OSD/Z). \quad (3)$$

The ground truth $(\theta gt)$ gaze angles for the AOI locations with coordinates (AOI_X, AOI_Y) were:

$$OSD\_GT = \mu\sqrt{(AOI\_X)^2 + (AOI\_Y)^2},$$
$$GT(\theta gt) = \tan^{-1}(OSD\_GT/Z). \quad (4)$$

The gaze yaw and pitch angles were derived as follows:

$$Gaze\ pitch(\theta pitch) = \tan^{-1}(GazeY/Z),\ Gaze\ yaw\ (\theta yaw) = \tan^{-1}(GazeX/Z). \quad (5)$$

The ground truth pitch and yaw angular values for each AOI dot with screen coordinates *(AOI_X, AOI_Y)* were given by:

$$AOI\ pitch = \tan^{-1}(AOIy/Z),\ AOI\ yaw = \tan^{-1}(AOIx/Z). \quad (6)$$

These gaze variables along with their statistics were later used to construct the feature vectors for the learning algorithms. The plots of the gaze frontal and rotational angles are shown below in Figure 4. Throughout this paper, the terminologies gaze angle, gaze yaw, and gaze pitch angle will be used to indicate the variables defined by Equations (3)–(6) above..

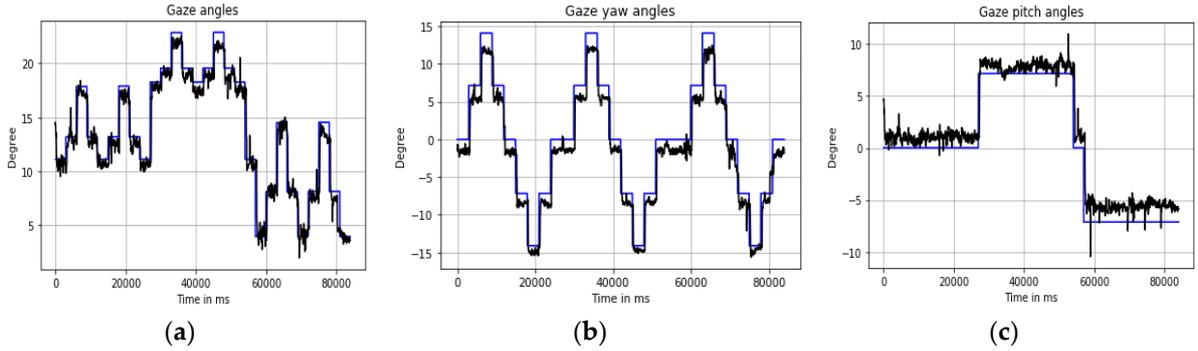

**Figure 4.** Time series of (**a**) gaze angles ,(**b**) gaze yaw, and (**c**) gaze pitch angles (black lines with ground truth in blue lines) for one person during one experimental session for a neutral pose captured on the desktop setup.

The common problems associated with analyzing the collected gaze dataset were: (a) The presence of outliers, (b) missing or null values, and (c) unequal data row lengths. Therefore, before using the data to train the models and yield meaningful results, it was essential that the data went through certain pre-processing steps. To fill the missing values, a mean substitution was used. For outlier removal, the following methods were tested: (i) 1-D Median filtering: This method can detect isolated out-of-range values from legitimate data features. In this method, the value of a data point is replaced by that of the median of all data points in a neighborhood w [14], such that:

$$y[m,n] = median\ \{x[i,j], (i,j) \in w\}. \quad (7)$$

(ii) Median absolute deviation (MAD): This is calculated by taking the absolute difference between each point and the median, and then calculating the median of those differences. This is more robust than using the standard deviation for outlier detection as the standard deviation is itself affected by the presence of outliers [15]. (d) Inter-quartile range(IQR): The concepts of using the Z-score and inter-quartile range (IQR) to study outliers were discussed in our previous study [1]. A data point is denoted as an outlier if the value for the point is 1.5·IQR above the third quartile or below the first quartile of the data. The results from the three outlier removal methods are shown in Figure 5a–c and it is seen from the figures that median filtering (with a kernel size =41, which is the mean number of data points around each AOI) worked best for all of the datasets.

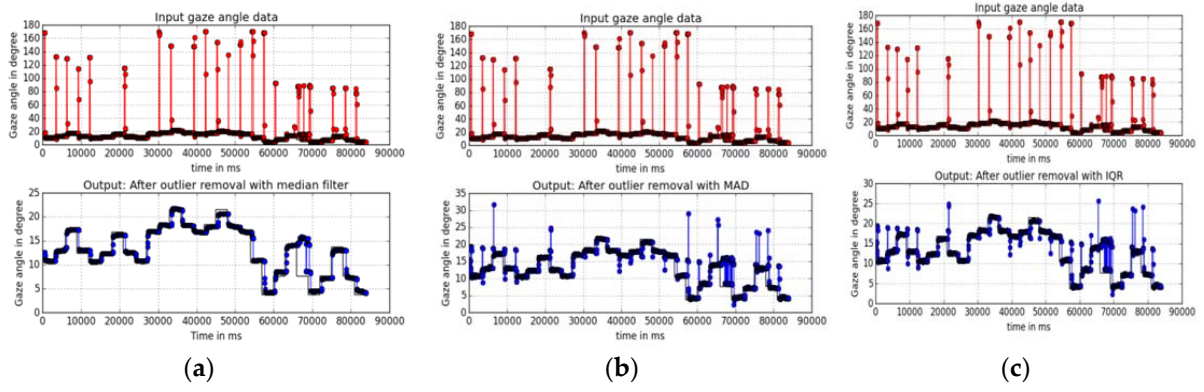

**Figure 5.** Outlier detection and removal with (**a**) median filtering, (**b**) MAD, and (**c**) IQR methods. Median filtering (**a**) is seen to remove nearly all outliers.

*2.5. Exploratory data analysis and visualizations*

Since the eye tracking datasets were collected under unconstrained conditions and the nature of such data or its distributions were unknown, data exploration was essential to observe any underlying patterns before proceeding to the machine learning step. This was done after the data cleaning steps described above as the presence of outliers and null values made it impossible to study any meaningful patterns prior to data preprocessing. After null and outlier removal, gaze angular errors for the frontal and rotational components were computed in absolute units (degrees) as the angular deviation between ground truth and estimated gaze angular values using Equations (3)–(6), as:

$$(Gaze\ frontal\ angular\ error)_i = (\theta gaze)_i - (\theta gt)_i, \qquad (8)$$

$$(Gaze\ yaw\ angle\ error)_i = (\theta yaw)_i - (AOI\_Yaw)_i, \qquad (9)$$

$$(Gaze\ pitch\ angle\ error)_i = (\theta gaze)_i - (AOI\_Pitch)_i. \qquad (10)$$

The gaze frontal angular error and gaze yaw and pitch errors are three categories derived from the same gaze data sample. This aspect was used to generate the training sample set from the collected gaze datasets.

2.5.1. Studying Gaze Data Statistics for Desktop and Tablets

The first step in studying the error characteristics of gaze data obtained from the different experiments was to look into the error statistical parameters, such as the mean, median absolute deviation, inter-quartile range, and 95% confidence interval. These parameters also form parts of the feature vector for the classification studies described later in this paper. Table 1 shows the statistical properties of the gaze errors for different desktop experiments. It was seen that the gaze error is higher at low user distances and the error reduces as the user-tracker distance increases. This was primarily due to a reduction in the visual angle and eccentricity with increasing distance as discussed in [1]. Error due to head yaw was seen to have the highest magnitude, although errors due to head pitch had the highest inter-quartile range or highest variability in error values. Additionally, the error levels due to various head poses were quite high compared to when the head pose was neutral (UD60 values in Table 1). All values in the tables units of degrees of angular resolution.

Table 2 shows the statistical properties of the gaze errors for different experiments on the tablet platform, including variations in the user distance and tablet pose (here the roll, pitch, and yaw represent tablet poses). The magnitudes of errors due to tablet pose changes were high and the highest error was caused due to platform roll variations. It was also seen that the error characteristics of tablet data are quite different than those from the desktop platform. Compared to desktop data, the error magnitudes were lower for the tablet for all user distances. However, the magnitude of errors produced due to different platform poses was higher compared to errors induced due to the head pose.

Table 1. Gaze error statistics from the desktop experiments.

| Desktop | UD50 | UD60 | UD70 | UD80 | Roll 20 | Yaw 20 | Pitch 20 |
|---|---|---|---|---|---|---|---|
| Mean | 3.37 | 2.04 | 1.21 | 1.02 | 3.7 | 8.51 | 3.15 |
| MAD | 3.49 | 1.77 | 0.82 | 0.66 | 3.63 | 10.0 | 1.90 |
| IQR | 1.13 | 0.77 | 0.76 | 0.79 | 1.21 | 1.49 | 1.59 |
| 95% interval | 3.15–3.59 | 1.90–2.18 | 1.15–1.26 | 1.16–1.24 | 3.30–4.09 | 7.60–9.43 | 2.83–3.47 |

Table 2. Gaze error statistics from the tablet experiments.

| Tablet | UD50 | UD60 | UD70 | UD80 | Roll 20 | Yaw 20 | Pitch 20 |
|---|---|---|---|---|---|---|---|
| Mean | 2.68 | 2.46 | 0.59 | 1.55 | 7.74 | 4.25 | 2.45 |
| MAD | 0.38 | 0.42 | 0.29 | 0.24 | 0.77 | 0.60 | 0.46 |
| IQR | 0.39 | 0.54 | 0.33 | 0.22 | 0.75 | 0.53 | 0.23 |
| 95% interval | 2.65–2.71 | 2.43–2.48 | 0.57–0.61 | 1.53–1.57 | 7.69–7.80 | 4.22–4.29 | 2.41–2.49 |

2.5.2. Studying Gaze Error Distributions for Desktop and Tablet Data

The one-dimensional distributions of the angular error values for different gaze datasets were studied using the kernel density estimate (KDE) [16]. Since the exact distributions of the gaze errors were unknown, the kernel density estimation was useful since it is a non-parametric way to approximate the probability density function of the data, compared to parametric estimation, where a fixed functional form and its parameters are required to fit the data. For data with samples x(i), using a kernel function ($K$) and bandwidth h, the probability density at a point x is:

$$KDE = \sum_{i=1}^{N} K(y - x_i)/h. \qquad (11)$$

The Gaussian kernel K is given by:

$$K(u) = \frac{1}{\sqrt{2\pi}} e^{-0.5u^2}. \qquad (12)$$

Using the data collected from the desktop and tablet experiments, the KDE plots of gaze errors for three different user distances were plotted (bandwidth = 0.2) in Figures 6 and 7 above. It was seen that different user distances have individual impacts on the gaze error distribution. These error patterns were difficult to decipher when looking at raw gaze data or simple error magnitudes. Further, the KDE plots were found to be non-Gaussian or resembling any known statistical distribution, which makes it difficult to predict nature of the gaze errors. These form the background for studying gaze error characteristics and implementing the learning tasks described in this paper.

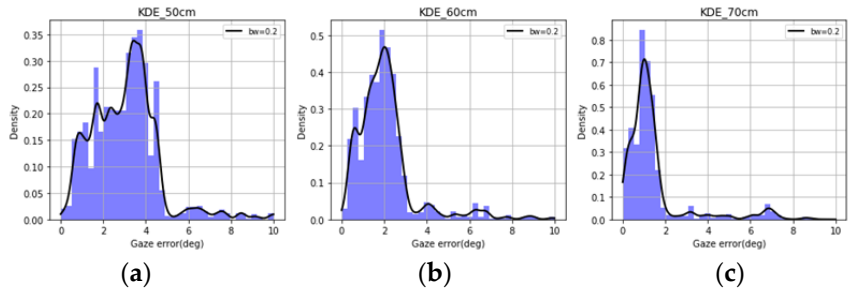

Figure 6. Kernel density plots over histograms of gaze error for user distances of (a) 50cm, (b) 60cm, and (c) 70cm for desktop data.

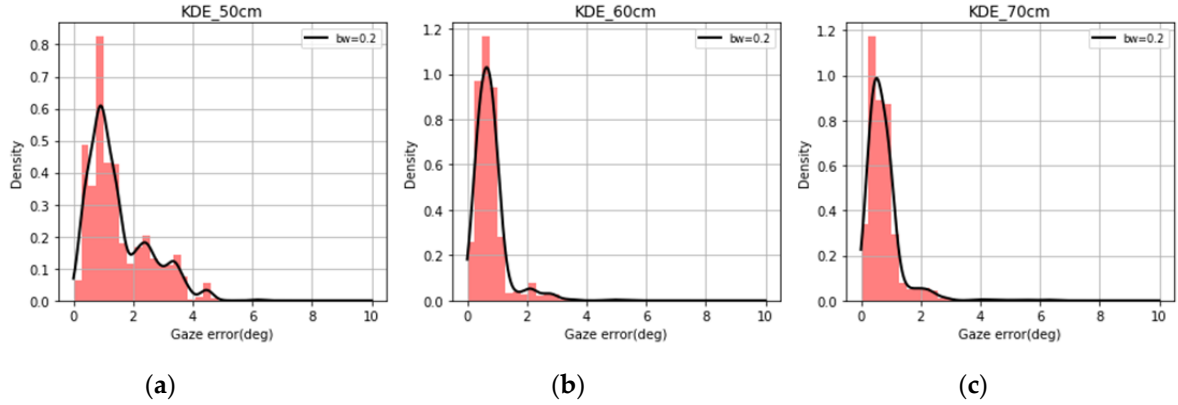

**Figure 7.** KDE plots over histograms of gaze error for user distances (**a**) 50cm, (**b**) 60cm, and (**c**) 70cm for tablet data.

2.5.3. Studying Spatial Error Distribution Properties

The two-dimensional spatial distribution of gaze error values over the display screen area, as a function of the corresponding visual angles for different datasets, were computed and are shown in Figure 8a–h. These plots show that factors like head pose angles have distinct impacts on the gaze error spatial patterns, with minimum errors obtained for neutral head positions. Similar features were seen from the tablet data plots, for different platform poses.

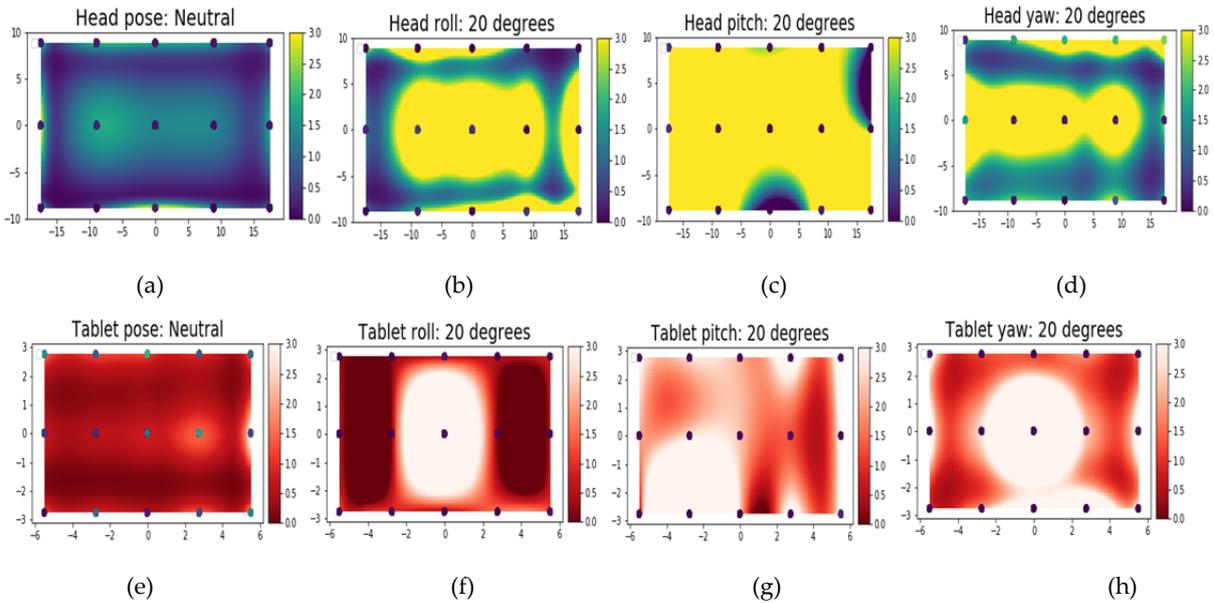

**Figure 8.** Gaze error spatial distribution as a function of the visual angles (x axis= yaw angle, y axis= pitch angle) over the display area. Figure 8 (**a**–**d**) show the spatial error variation due to the head pose (desktop). Figure 8 (**e**–**h**) show the error distributions due to the tablet poses.

2.5.4. Studying Data Correlations

Desktop data from user distance experiments at 50, 60, 70, and 80 cm (UD 50, UD60, UD70, UD80) and head pose experiments at head roll, pitch, and yaw angles of 20 degrees, (R20, Y20, P20) were used to compute the desktop data correlation matrix of Figure 9a. Similarly, data from the tablet experiments for four user distances and three platform poses were used to compute the tablet data correlation matrix of Figure 9b. As can be observed, the gaze data collected under different operating conditions from the same platform and eye tracker do not have any correlations between their characteristics.

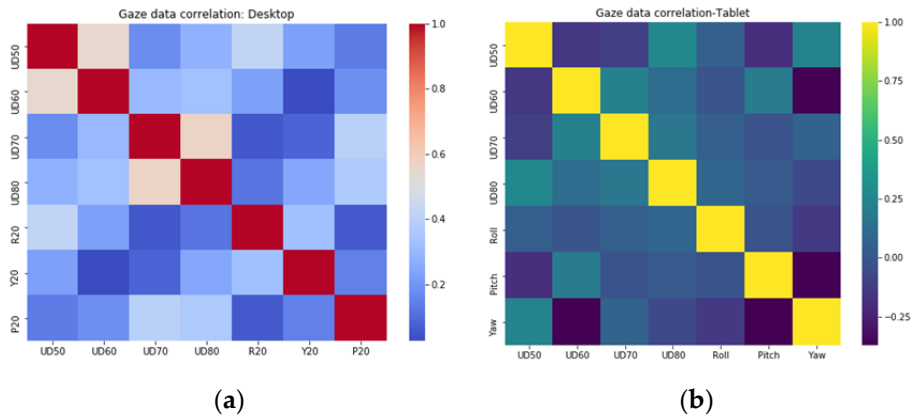

|  (**a**) | (**b**) |

**Figure 9.** Correlation between data from (**a**) desktop and (**b**) tablet experiments.

2.5.5. Discussions

In the above sections, various parts of the gaze data processing pipeline were described, and detailed visual and numerical exploration of the eye tracking data was done. With respect to the data demographics, no age bias could be found, and the effect of glasses were ruled out by having no participants wearing glasses during the experiments. There was also no gender bias found in the data. The chin rest had to be adjusted according to individual participants' heights to make the participants head level with the center of the display screen and so that their eyes could be clearly visible to the eye tracker's camera. Participants were made familiar with the experiments through instructions and pilot experiments.

The gaze data showed varied levels of in-homogeneities, and only after outlier removal could distinct gaze error patterns be observed. One significant aspect noted is that the gaze errors from the tablet are much lower than the errors from the desktop for the same user distances. This could indicate that the distinguishing aspect between the two platforms, which is the display size and resolution, could be a factor determining the error levels for an eye tracker. However, further comparison of the data from the two platforms was not done in this study. Other takeaways from this section include the identification of robust outlier removal methods for gaze data and observations of the gaze error distributions, which were heavily affected by operating conditions but were not observable in the raw gaze data plots. Additionally, studies on the correlation of different eye tracking datasets (Figure 9) revealed an important aspect: Under different operating conditions, an eye tracker's data may behave in totally independent ways, which are not related to the tracker's data characteristics under stable conditions.

In order for other researchers to compare their eye trackers' data characteristics with the data collected during this study, the full set of data collected for the different setups and operating conditions was provided in the Mendeley open data repository and can be accessed in the link here: https://data.mendeley.com/datasets/cfm4d9y7bh/1 (See Supplementary materials section at the end of this paper)

## 3. Identification of Error Patterns in Eye Gaze Data

*3.1. Objectives and Task Definition for Classification of Desktop and Tablet Data*

In this section, the main goal was to identify a certain error source or operating condition solely from the output data from an eye tracker, which was influenced by the condition. For this purpose, multiple classifier models were trained using data collected from the different eye tracking experiments, which were seen to produce different error patterns. The objective was to see if machine learning models [17,18] could learn to distinguish between these gaze error patterns as they appear among a mix of data captured under different operating conditions.

The following classification tasks were performed using the desktop datasets: (1) Classification of errors for different user distances (i.e., between four classes of data from a user distance of 50-, 60-, 70-, 80-cm datasets), (2) classification of errors for different head poses (i.e., between four classes: neutral pose, roll (20 degrees), pitch (20 degrees), and yaw (20 degrees) datasets), and (3) classification between the head pose and user distance errors patterns (i.e., between four user distance classes, and three head pose classes, with a total of seven classes). With data from the tablet experiments, the following classification tasks were implemented: (1) Classification of errors for different user distances (i.e., between four classes of user distances data), (2) error classification for different tablet orientation poses (i.e., between four classes of data from the neutral tablet pose, roll (20 degrees), pitch (20 degrees), and yaw (20 degrees) datasets), and (3) classification between the tablet orientation and user distance error patterns (i.e., between four user distance classes, and three tablet pose classes, with a total of seven classes).

Before implementing the classification algorithms on both the desktop or tablet data, the user distance, head-pose, and tablet orientation datasets were augmented to increase the number of training samples [19,20]. After this, training features were constructed and formatted before being input into the models. Same data augmentation strategies were used for both the desktop and tablet data. For the training and testing, the gaze angle, yaw, and pitch feature datasets were created and used. The classification results from the desktop data are provided in Section 3.3 while that from the the tablet datasets are in Section 3.4.

*3.2. Data Augmentation Strategies*

With our dataset size (20 persons x 4 operating conditions each for the user distance and head pose), in order to use a sufficient number of features for classification without facing the overfitting problem, augmentation of the dataset was essential. In this study, 10-fold augmentation strategies were used on the gaze angle, yaw, and pitch error datasets estimated from raw data using Equations (8)–(10). The methods used for data augmentation were as follows: 1) The addition of Gaussian noise: Gaze error magnitudes at all data points were perturbed with Gaussian noise with 0 mean and 0.2 sigma [21]; (2) addition of jitter: Human eye jitter was modelled as pink noise [22,23], in which the power spectral density is inversely proportional to the frequency of the signal, given by PSD= $1/f^{\alpha}$. The jitter signal was simulated by first generating white noise with the mean, sigma (0, 0.2), and successively applying a pink noise filter with the parameters $\alpha = 0.8$ at a frequency of 2Hz and added to the error datasets; (3) interpolation: Linear interpolation on the input data points was used to produce variants of the original gaze error samples [24];(4) convolution: The error signals were convolved with a raised cosine kernel with a window size of N = 30 with the form of Equation (13) below to produce a smoothed variant of the original data [25]:

$$w(n) = 1/2\left(1 - \cos(\frac{2\pi n}{N-1})\right). \tag{13}$$

The convolution operation was given by:

$$(f * g)(t) = \int_{-\infty}^{\infty} f(t-\tau)g(\tau), \tag{14}$$

(5) time shifting: The gaze data were shifted by 10 samples to estimate new variants in error values from the shifted datasets [26]; (6) combinations: Combinations of the addition of different noise patterns to interpolated signal were used to augment the dataset; and (7) flipping: Horizontal and vertical flipping of error magnitudes at the different AOIs were used to augment the dataset as well as to remove any viewer bias towards the top, bottom, and side locations of the screen. In horizontal flipping, the error magnitudes of the top AOI-1 to AOI 5 (Figure 2b) were replaced by the values of the bottom AOIs (AOI-11 to AOI 15). In vertical flipping, the error magnitudes of the left AOIs (AOI No. 1,6,11) were swapped with the right AOIs, (No. 5,10,15). Figure 10 shows how the samples from a dataset were modified by these augmentation methods. Table 3 shows how training samples were created from the collected dataset.

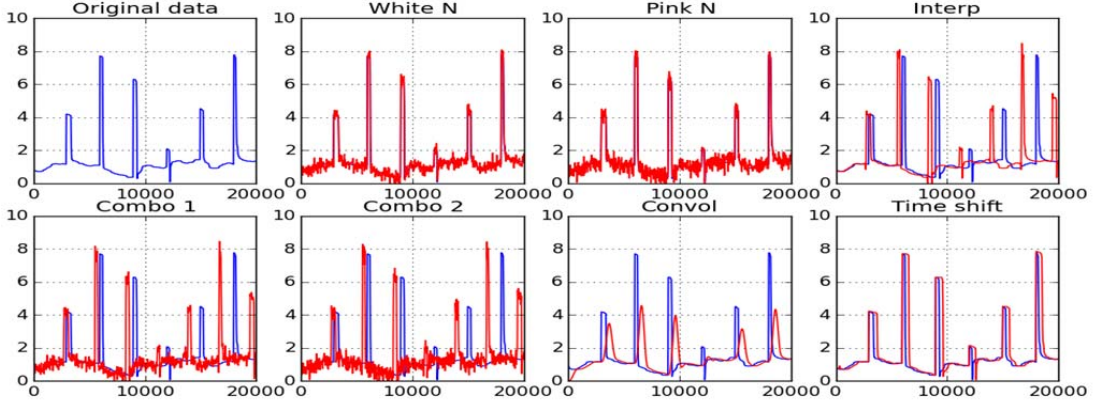

**Figure 10.** Shows a sample from a dataset after applying each augmentation strategy. The blue lines represent samples from the original data, while the red lines are from the augmented dataset after applying each strategy.

*3.3. Feature Engineering, Exploration, and Selection*

The original data sample set in this study had 20 participants and for each participant, there were three categories, i.e., gaze angular error, yaw error, and pitch error (Equations (8)–(10)) from which training features were computed [27–30]. The training feature set was constructed by estimating the error magnitudes at 15 AOI locations distributed all over the display screen (to account for the spatial distribution of errors), and statistical values for each sample, i.e., mean error ($\mu$), error standard deviation ($\sigma$), interquartile range (IQR), and upper and lower bounds of the 95% confidence interval of the sample. Thus, each training sample had 20 features as follows:

[Error_AOI-1, Error_AOI-2,.Error_AOI-15, $\mu$, $\sigma$, IQR, 95% conf upper, 95% conf lower]$_{sample.}$    (15)

The above feature set was calculated from the gaze angle as well as the yaw and pitch angle data for each sample. To train the machine learning models, the features from all datasets were standardized so that they had a zero mean and unit variance and were shuffled randomly before splitting the datasets into the train and test sections. The proportion of train vs. test samples was varied between 0.4 to 0.25. Table 3 below describes the contents of all the training and test datasets used in this study.

For visualization of the high-dimensional feature set, the t-distributed stochastic neighbor embedding (t-SNE) method [31] was used, which maps data points $x_i$ in the high-dimensional feature space $R^D$ (D is the dimension of the feature set, D=20 here) to points $y_i$ in a lower d-dimensional space ($R^d$, here d= 2) by finding similarities ($p_{ji}$) between the data and learning the corresponding low dimensional mapping points $y_1$, …, $y_N$ (with $y_i \epsilon R^d$ that reflects these similarities as best as possible [32–34]. The pairwise similarity ($p_{ji}$) between points $x_i$ and $x_j$ is:

$$p_{ji} = = \frac{\exp(-\|x_i - x_j\|^2 / 2\sigma_i^2)}{\sum_{k \neq i} \exp(-\|x_i - x_k\|^2 / 2\sigma_i^2)}.$$    (16)

To observe the training data in the feature space, the t-SNE algorithm (with n_components = 2, perplexity = 80) was applied on the feature sets computed from the user distance, head pose, and platform pose datasets, as plotted in Figure 11(a) and Figure 11(b) for the desktop and tablet data, respectively. On the datasets from both the desktop and tablet, the t-SNE points look scattered with no structure. For the head pose data and platform pose data and mixed datasets, the t-SNE plots show several clusters, but within them, the class labels were found to be highly mixed. This reflects the high degree of complexity of the datasets used in this study since the different classes within them do not form any observable clusters; they are neither symmetrically distributed nor have any clear separation between them. This is close to real world gaze datasets in which normal gaze data is most often mixed with anomalous data due to unpredictable operating conditions.

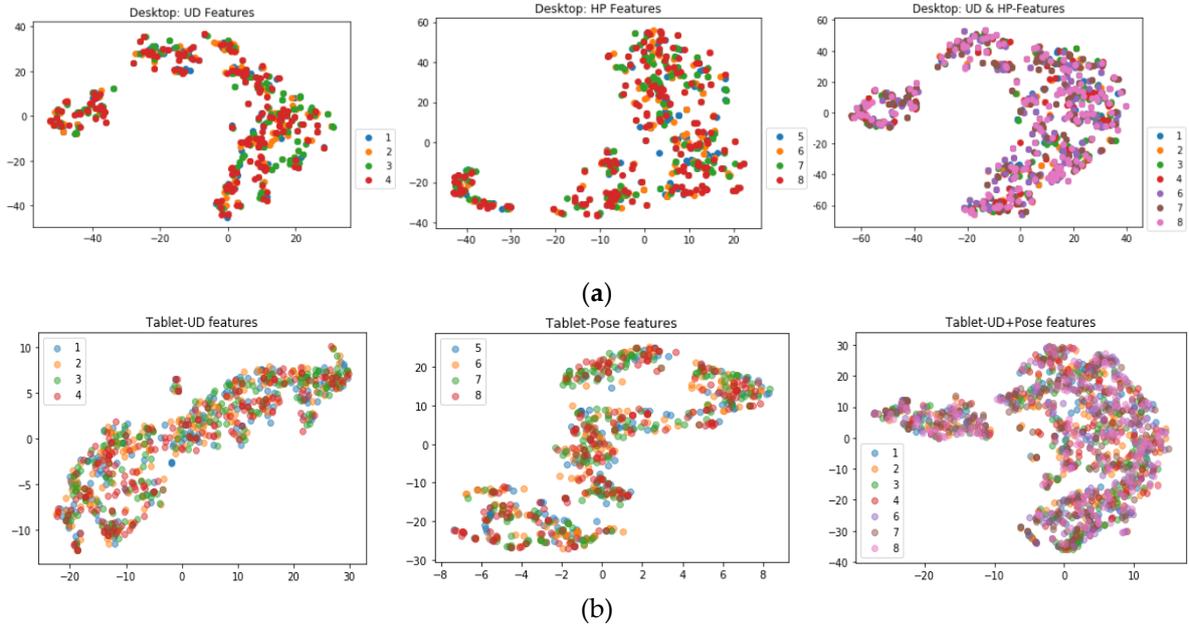

**Figure 11.** (**a**). t-SNE plots for (i) user distance (ii) head pose (iii) merged user-distance and head pose for desktop datasets. Legends 1, 2, 3, and 4 are user distance classes UD50-UD80, 5, 6, 7, and 8 are head pose classes the neutral, roll, pitch, and yaw. (**b**). t-SNE plots for (a) the user distance, (b) tablet pose, and (c) merged user-distance and tablet pose for tablet datasets. Legends 1, 2, 3, and 4 are user-tablet distance classes UD50-UD80, 5, 6, 7, and 8 are tablet pose classes neutral, roll, pitch, and yaw.

**Table 3.** Training and test dataset details.

| Samples for Train, Test | Original Feature Set (20 Features) | Augmentation Strategies | Samples Per Person | Samples and Classes |
|---|---|---|---|---|
| **Total participants: 20** 12- 16 participants for training, 8-5 for test Data labelled and randomly shuffled | (1) Gaze error values at 15 AOIs, Mean, SD, IQR, 0.95 interval bounds of gaze error (2) Yaw error values at 15 AOIs, Mean, SD, IQR, 0.95 interval bounds of yaw error (3) Pitch error values at 15 AOIs, Mean, SD, IQR, 0.95 interval limits of pitch error **Reduced feature set (5** | (1) Gaussian noise (2) Jitter or pink noise (3) Horizontal data flipping (4) Vertical data flipping (5) Magnitude warping (6) Time warp (7) Interpolation and combinations | 10 x gaze error 10 x yaw error 10 x pitch error **Merged dataset** 30 samples | (1) Desktop user dist, 2400 samples, 4 classes (2) Head pose, 2400 samples, 4 classes (3) Desktop mixed: 4200 samples, 7 classes (5) Tablet user dist, 2400 samples, 4 |

| | | |
|---|---|---|
| **features)**<br>Mean, SD, IQR, 0.95 interval bounds for each sample. | | classes |
| | (6) | Tablet pose, 2400 samples, 4 classes |
| | (7) | Tablet mixed: 4200 samples, 7 classes |

For feature selection, the relative importance of features (Equation (15)) were estimated using a random forest classifier method [35–37]. This model comprises of a number of decision trees in which it can be computed how much each feature reduces the weighted impurity [38] in the tree. The total impurity decrease due to each feature was averaged to rank the feature's importance.

Features computed on desktop and tablet datasets were fed to a random tree classifier model with hyperparameters (n_estimators = 200, max_depth = 8). The rankings of features for the desktop and tablet datasets are shown in Figure 12a,b. In all the datasets, the mean, standard deviation, IQR, and confidence intervals (feature numbers 16–20) emerged as the most significant features. Based on this, the following reduced feature set was computed: [μ, σ, 95%Conf_up, 95%Conf_down, IQR]. This reduced feature set was used with SVM and KNN for classification. However, the neural network models required the full feature set to reduce the training error and prevent under-fitting.

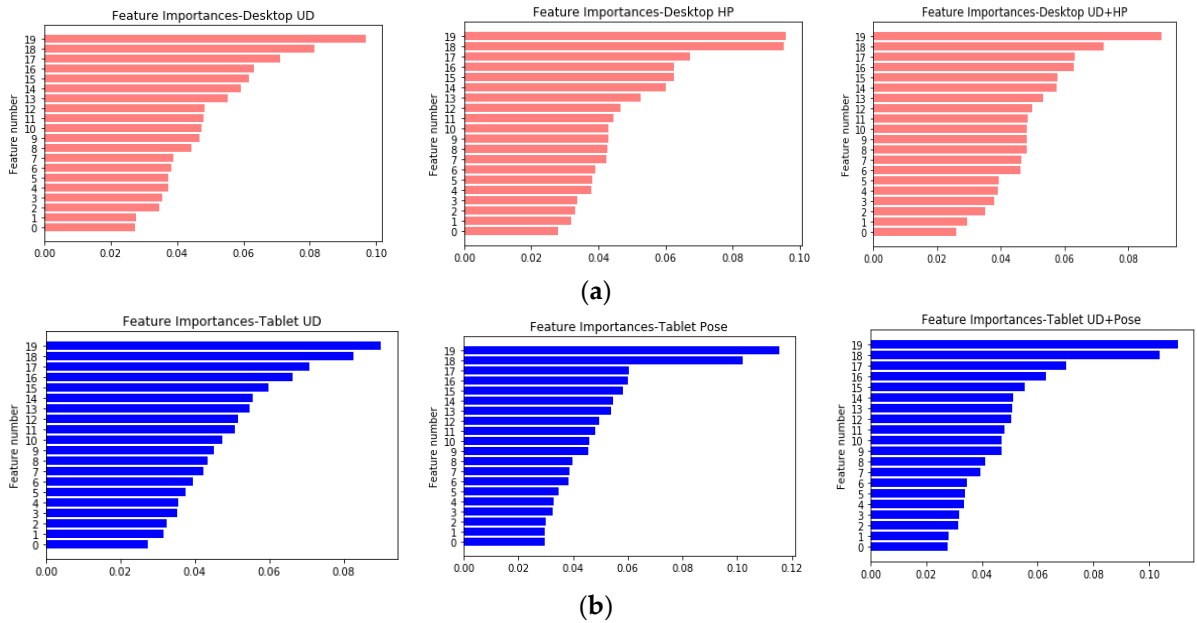

**Figure 12.** (**a**): Relative importance of features for different desktop datasets. (**b**): Relative importance of features for different tablet datasets.

*3.4. Classification Models: k-NN, SVM, and ANN*

After data augmentation and exploration of the feature set and labelling, the datasets were used with three different machine learning models from the Python Scikit-Learn libraries, run on a Windows 7 computer with core i7 2.6 GHz processor. A brief description of the models is given below.

(a) K-nearest neighbors (k-NN): In this, the underlying assumption is that samples of the same class will be the nearest neighbors of each other, i.e., the distance between them will be small since they are related [39,40]. While using this method, the training samples form vectors in a multidimensional feature space, each with its class label. In the training phase of the algorithm, the feature vectors are stored along with class labels of the training samples. In the classification phase, the distance between an unlabelled input vector and the training dataset is calculated to get the k (a user defined hyperparameter number) nearest points of the input sample. Then, the input sample is categorized into the class of the majority in the k nearest points. Thus, if the training data is {($x_1, y_1$), ($x_2, y_2$),…, ($x_N, y_N$)} and x is the feature vector of an input sample, the KNN finds the class of the input sample x using a distance metric [41,42], which in this study is the Euclidean distance given by:

$$Dist(x_i, x_j) = \sqrt{(\sum_{i=1}^{N}(x_i - x_j))^2} \ . \tag{17}$$

(b) Support vector machines (SVMs): It is a supervised learning method [43,44] that works on the principle of transforming the input feature space by a nonlinear transformation to a high-dimensional feature space, and searching for an optimal separating hyperplane for the input classes in this new high-dimensional space [45,46]. With this separating hyperplane, the training data xi with labels yi can be classified so that the minimal distance of each point from the hyperplane is maximized. The training data are depicted as an instance-label pair (xi, yi), i = 1,…,m, where xi ϵ Rn represents the input vector and yi ϵ (−1,1) is the corresponding output label of xi. The objective function for SVMs can be defined as:

$$\min_{w,b,\xi} \frac{1}{2}\|w\|^2 + C\sum_{i=1}^{N}\zeta^2, \tag{18}$$

with the condition:

$$y_i(w^T x + b) \geq 1 - \xi^2 \ (i=1,…N). \tag{19}$$

where C >0 is the regularization parameter and $\xi_i$ is the slack variable, w is the weight vector, and b is the offset. To classify unlabelled examples $x_k$ according to labels $y_k$ using the kernel function K($x_i$, $x_k$), the optimal separating hyperplane equation becomes:

$$y_k = sign(\sum_{x_i} a_i y_i K(x_i, x_k) + b, \tag{20}$$

where S is the set of support vectors xi, and ai are Lagrange multipliers (used for solving the optimization problem). In this study, a Gaussian radial basis function or RBF kernel is used, which is given by:

$$K(x_i, x_k) = \exp\left[-\gamma \|x_i - x_k\|^2\right]. \tag{21}$$

The hyperparameters C and $\gamma$ are chosen for optimal fitting and discussed in the next section.

(c) Multilayer perceptrons (MLP) or neural networks: There are supervised learning algorithms [47–50], which takes in labelled training examples and solves the complex non-linear hypothesis by a network of computing units called "neurons". Learning in an MLP takes place by updating the connection weights of each neuron after passing a batch of data samples from input to output, depending on the amount of error in the output compared to the target. The general update rule for weights ($\Delta w_{ji}$) in an MLP based on backpropagation and the gradient descent is:

$$\Delta w_{ij} = -\eta \frac{\partial E(n)}{\partial w_{ij}} y_i(n), \tag{22}$$

where $y_i(n)$ is the output of the previous neuron and η is the learning rate, and E is the error in the nth node. In this study, neural networks with ReLU activation, Adam optimizer, and a constant learning rate of 0.001 were used.

*3.5. KNN, SVM, and MLP-Based Classification Results on Desktop Data*

In the sub-sections below, the results from training the three ML models with the dataset described in Table 3 are presented, along with statistics of the learning process. The models used here were optimized through experimentation to select the set of hyperparameters that best improve their cross-validation scores. In some cases, the results show underfitting, especially in seven class classification problems, which are inherently complex. For this kind of problem with little separation between classes, model underfitting is a commonly observed issue.

3.5.1. Results, Task I: Classification of Desktop Data for Different User Distances

The KNN, SVM, and MLP models were used for the classification of gaze datasets corresponding to four different classes of user distances (50, 60, 70, 80 cm) and the results are given in Table 4. In the table, only the cross-validation accuracies are mentioned. A training-test sample proportion of 30% and 10-fold cross validation was used in all cases [51]. For all models, grid search [52] with cross-validation was used to determine the optimal hyperparameters. For K-NN, the optimal number of neighbors was found to be 3. For SVM, the RBF kernel was used [53], with the soft margin cost function parameter C set to 10 and gamma set to 1.0. C defines the trade-off between misclassification and the simplicity of the decision surface. Gamma is the RBF kernel parameter (Equation (21)). For MLP, the grid search yielded a best hidden layer size of 3 with neuron configurations of [50, 100, 50] units in the layers, with a full training set of 20 features being used.

3.5.2. Results, Task II: Classification of Desktop Data for Different Head Poses

In this task, the gaze datasets collected from the desktop setup for different user head poses were used. The head pose datasets include the gaze error values for the head roll (20 degrees), pitch (20 degrees), yaw (20 degrees), and neutral (roll, pitch, yaw = 0 degrees). The KNN, SVM, and MLP models were used and the classification accuracies for the different classifiers are tabulated in Table 5.

Using the grid search, for KNN, the best number of neighbors was found to be 3 and a reduced feature set was used. For SVM, a C value of 10 and gamma of 1.25 was used with the reduced feature set. For the MLP model, a three-layer network was used, and its architecture was tuned by varying the number of units in each layer between 50 and 100 and varying the regularization parameter values between 0.001 to 0.5 to control overfitting. Additionally, the full feature set was used for training.

3.5.3. Results, Task III: Classification on Merged User Distance and Head Pose Datasets from Desktop

In this task, the head-pose and user distance datasets were merged, and classifiers were applied to do a seven-class classification on this mixed dataset. The classification results are presented in Table 4 and the confusion matrix for the MLP model is shown in Figure 13. Since the head pose data was collected at a user distance of 60 cm, the neutral head pose data and the UD60 data are the same, and therefore only one of these two datasets were used to avoid a class imbalance in the mixed dataset. As above, the three classifiers were trained and tested on this dataset. For the KNN, the best number of neighbors was chosen to be 3 as shown in Figure 14a below, which shows the dependence of the train, test, and cross-validation error as a function of the number of neighbors used in the model. For the SVM model, the parameters were the same for the above two tasks. For the MLP, 2 hidden layers with 100 units each and a regularization value of 0.001 were used with the full set of features. Table 5 shows the performance of the three classifiers on different datasets, with true and false detection rates and precision values. It is observed that for desktop data, the KNN and MLP classification performances are close, with KNN performing well for all the datasets.

Table 4. Classifier accuracies for different datasets and machine learning models.

| Dataset | For user Distance | | | For Head Pose | | | On Mixed Datasets | | |
|---|---|---|---|---|---|---|---|---|---|
| | k-NN | SVM | MLP | k-NN | SVM | MLP | k-NN | SVM | MLP |
| **Only gaze angle features** | 89% | 69% | 86% | 91% | 83% | 90% | 87% | 74% | 83% |
| **Only gaze yaw features** | 83% | 64% | 84% | 88% | 78% | 86% | 81% | 66% | 80% |
| **Only gaze pitch features** | 96% | 89% | 73% | 95% | 96% | 83% | 92% | 91% | 79% |
| **Merged (gaze, yaw, pitch) feature dataset** | 84% | 67% | 81% | 89% | 80% | 84% | 85% | 70% | 77% |

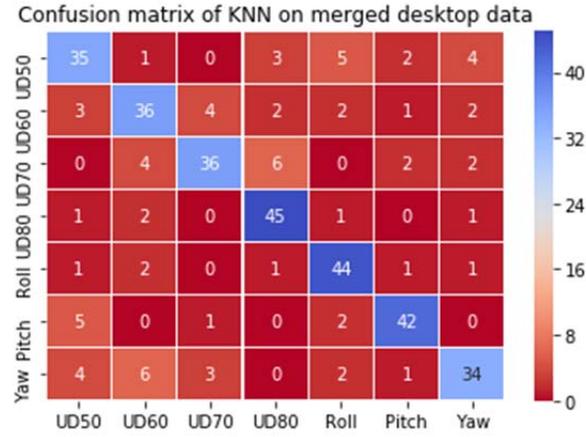

**Figure 13.** Confusion matrix using K-neared neighbor on the mixed head pose and user distance dataset for the desktop.

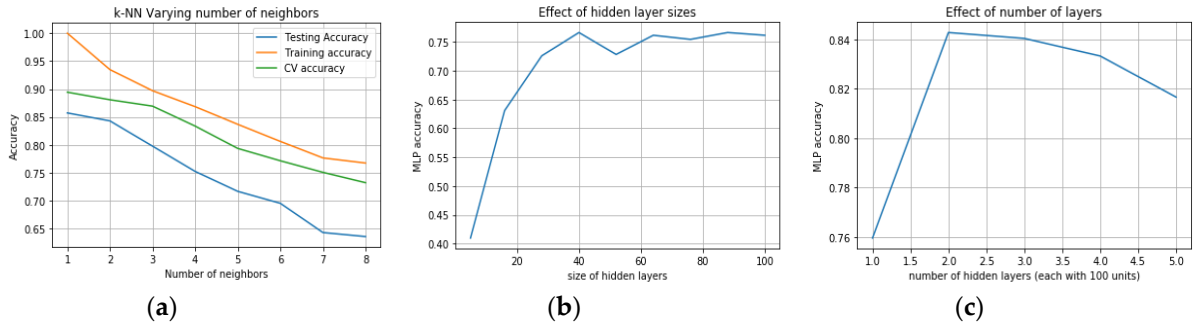

(**a**)             (**b**)            (**c**)

**Figure 14.** Effect of varying the model hyperparameters for KNN (**a**) and (**b**,**c**) Multilayer perceptron

**Table 5.** True and false detection rates from classifiers of the desktop (user distance, head pose, and mixed datasets).

| For User Distance Dataset | | | | For Head Pose Dataset | | | | On Mixed Dataset | | | |
|---|---|---|---|---|---|---|---|---|---|---|---|
| | KNN | SVM | MLP | | KNN | SVM | MLP | | KNN | SVM | MLP |
| **TPR** | 0.96 | 0.95 | 0.85 | **TPR** | 0.97 | 0.91 | 0.95 | **TPR** | 0.83 | 0.93 | 0.83 |
| **FPR** | 0.01 | 0.01 | 0.04 | **FPR** | 0.01 | 0.02 | 0.01 | **FPR** | 0.02 | 0.01 | 0.02 |
| **TNR** | 0.98 | 0.98 | 0.95 | **TNR** | 0.99 | 0.97 | 0.98 | **TNR** | 0.97 | 0.98 | 0.97 |
| **FNR** | 0.04 | 0.04 | 0.14 | **FNR** | 0.03 | 0.08 | 0.04 | **FNR** | 0.16 | 0.06 | 0.16 |
| **Precision** | 0.98 | 0.93 | 0.85 | **Precision** | 0.97 | 0.97 | 0.95 | **Precision** | 0.85 | 0.93 | 0.85 |

## 3.6. KNN, SVM, and MLP-Based Classification Results on Tablet Data

Here the results from using the tablet data with various classification models are presented. These include datasets for different user-tablet distances (50, 60, 70, 80 cm) and tablet poses of the neutral, platform roll, pitch, and yaw of 20 degrees. A mixed dataset was created by merging the datasets for the user distance and platform pose, and classifier models were trained to do seven class classifications, i.e., distinguish between four user distance and three tablet pose classes.

### 3.6.1. Results, Task IV: Classification of Tablet Data for Different User Distances

For the user distance dataset from the tablet platform, the KNN classifier with three neighbors was used. The SVM was used with the RBF kernel, C = 10 and gamma of 1, and the MLP classifier was used with 3 layers, 200 units each, and with a regularization value of 0.0001. The best performance was by MLP and the classification results are shown in Table 6.

Table 6. Classifier performance for different datasets.

| | For user-Tablet Distance | | | For Tablet Pose | | | On Mixed Datasets | | |
|---|---|---|---|---|---|---|---|---|---|
| Dataset | k-NN | SVM | MLP | k-NN | SVM | MLP | k-NN | SVM | MLP |
| **Only gaze angle features** | 81% | 82% | 81% | 86% | 88% | 75% | 80% | 80% | 78% |
| **Only gaze yaw features** | 84% | 86% | 78% | 93% | 92% | 85% | 84% | 84% | 70% |
| **Only gaze pitch features** | 90% | 81% | 71% | 94% | 91% | 89% | 84% | 84% | 79% |
| **Merged (gaze, yaw, pitch) feature dataset** | 85% | 84% | 76% | 88% | 85% | 82% | 83% | 83% | 75% |

### 3.6.2. Results, Task V: Classification of Tablet Data for Different Tablet Poses

For the tablet pose dataset, the KNN classifier with three neighbors was used. The SVM was used with RBF kernel, C = 5, and gamma of 0.5. The MLP model was used with 2 layers, 100 units each, and a regularization value of 0.0001. The best classification results were by KNN and are shwon in Table 6.

### 3.6.3. Results, Task VI: Classification of Tablet Data for Mixed User Distance and Tablet Pose Datasets

For the mixed user distance and platform pose datasets, the KNN classifier with three neighbors was used. The SVM was used with RBF kernel, C = 5, and gamma of 0.5 and the MLP classifier was used with 2 layers, 100 units each, and with a regularization value of 0.0001. Both SVM and MLP performed well and the classification results are shown in Table 6.

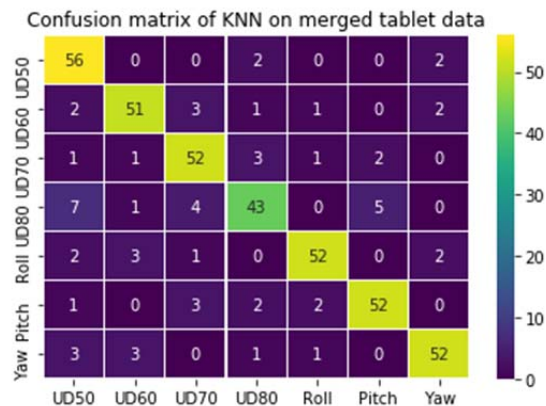

**Figure 14.(d)** Confusion matrix using K-nearest neighbour on the mixed head pose and user distance dataset for the tablet.

*3.7. Discussion*

The results demonstrate that ML models can distinguish between gaze data collected under normal and varying operating conditions. Thus, when using these models, anomalies present in gaze datasets may be detected. This is similar to the anomaly detection approaches [54–56] used in fields, such as cyber intrusion and video surveillance, where training sets of normal and nominal examples are used to design a decision rule such that occurrences of erroneous data samples are detected.

For the classification tasks, gaze datasets were constructed such that they contained signatures of a single or multiple error sources. It was found through the decision tree-based feature selection technique that statistical attributes, such as gaze error confidence levels and interquartile ranges, are significant parameters that can be used to distinguish gaze error sources. In the classification tasks, the KNNs and MLP classifiers performed the best for all datasets, although the MLP models take more time to converge. Additionally, it is seen from Tables 4 and 6 that using the gaze yaw and pitch features result in better detection accuracies than the frontal gaze datasets. From Tables 5 and 7, it is seen that the rate of false detections is also low for all the models. All the datasets, especially the ones created with seven classes, were found to be quite complex as no class separation could be detected from the t-SNE results. Despite this, the classification models achieved a cross-validation score of 85%–90% on most datasets. This shows the feasibility of the ML models in detecting gaze error sources even from complex gaze datasets where more than one source is present.

**Table 7.** True and false detection results from classifiers of the tablet (user distance, tablet pose, and mixed datasets). *Prec refers to the precision of classification.

| User-tablet distance dataset | | | | Tablet pose dataset | | | | Mixed dataset | | | |
|---|---|---|---|---|---|---|---|---|---|---|---|
| **UD** | **KNN** | **SVM** | **MLP** | **Pose** | **KNN** | **SVM** | **MLP** | **Mixed** | **KNN** | **SVM** | **MLP** |
| TPR | 0.87 | 0.85 | 0.71 | TPR | 0.90 | 0.83 | 0.79 | TPR | 0.80 | 0.83 | 0.70 |
| FPR | 0.04 | 0.04 | 0.09 | FPR | 0.03 | 0.27 | 0.06 | FPR | 0.03 | 0.02 | 0.04 |
| TNR | 0.95 | 0.95 | 0.90 | TNR | 0.96 | 0.97 | 0.93 | TNR | 0.96 | 0.97 | 0.95 |
| FNR | 0.12 | 0.14 | 0.28 | FNR | 0.09 | 0.16 | 0.20 | FNR | 0.19 | 0.16 | 0.29 |
| Prec.* | 0.87 | 0.88 | 0.72 | Prec.* | 0.91 | 0.86 | 0.79 | Prec.* | 0.82 | 0.88 | 0.72 |

**4. Modelling and Prediction of Gaze Errors**

Using the collected desktop and tablet datasets, regression models [57,58] were trained to predict gaze estimation errors using the gaze angle, yaw, and pitch angle values as input features. Two different models were built, which may be used to predict the gaze errors of an eye tracker under two different operating conditions. These include the head-pose error model, which can be used to predict gaze errors occurring due to various user head poses. The other is the platform pose error model, for predicting gaze error corresponding to different poses of the eye tracker. The gaze errors were predicted as a function of a user's gaze angle and gaze yaw/pitch angles (Equation (28)). The results from the various gaze error models are presented in sub-sections 4.1 and 4.2 below.

A regression model describes a dependent variable Y as a function of an independent variable x with the generic relation: Y = a + b1X1 + b2X2 + b3X3 + ... + btXt + u, where *X1, X2... Xt* are the independent variables, a is the intercept, b is the slope, and u is the regression residual [59]. The cost function used for the evaluation of the model fit is given by the root mean squared error or RMSE:

$$\text{RMSE} = \sqrt{\frac{1}{n}\sum_{i=1}^{n}(\hat{y}_i - y_i)^2}, \quad (23)$$

where n is the number of observations, $y_i$ is the true value of target to predict, and $\hat{y}_i$ is the model's predicted result. The optimization function of a standard linear regression can be expressed as:

$$\min ||Xw - y||^2, \quad (24)$$

where $X$ is the set of feature variables, $w$ represents the weights, and $y$ comprises of the ground truth data points. In this study, several regression models were used, including Ridge, Lasso, ElasticNet [60,61], and neural network-based models [62]. In Ridge regression, the issue of high variance is mitigated through the addition of a squared bias factor as regularization in the form:

$$\min ||Xw - y||^2 + z||w||^2. \quad (25)$$

Whereas in Lasso regression, an absolute value bias of the form below is used:

$$\min ||Xw - y||^2 + z||w||. \quad (26)$$

The ElasticNet regression uses the regularization factors of both the above techniques:

$$\min ||Xw - y||^2 + z\_1||w|| + z\_2||w||^2. \quad (27)$$

In this study, the regression models were used to map the gaze frontal and yaw, and pitch angle values to gaze errors produced under different operating conditions. The task of the regression algorithms is not only to find the mapping between input and output variables but also take into consideration the interactions between the input variables. The input features (X1, X2, X3) were:

$$[Gaze\_Angle, Gaze\_Yaw, Gaze\_Pitch] \rightarrow Gaze\ error. \quad (28)$$

For the regression tasks, the input features were standardized, such that their distributions had a mean value of 0 and standard deviation of 1. Six different regression models for predicting the gaze errors were trained on the features created using Equation (28) on the head pose (desktop) and platform pose (tablet) datasets. For each model, the fit estimates in RMSE values are summarized in Table 8. The coefficients of the best performing models are listed in Table 9.

*4.1. Head Pose Error Model*

The plots for the actual gaze error values due to the head pose (obtained from desktop setup) and the error values predicted after training by the different models are shown in Figure 15a–d.

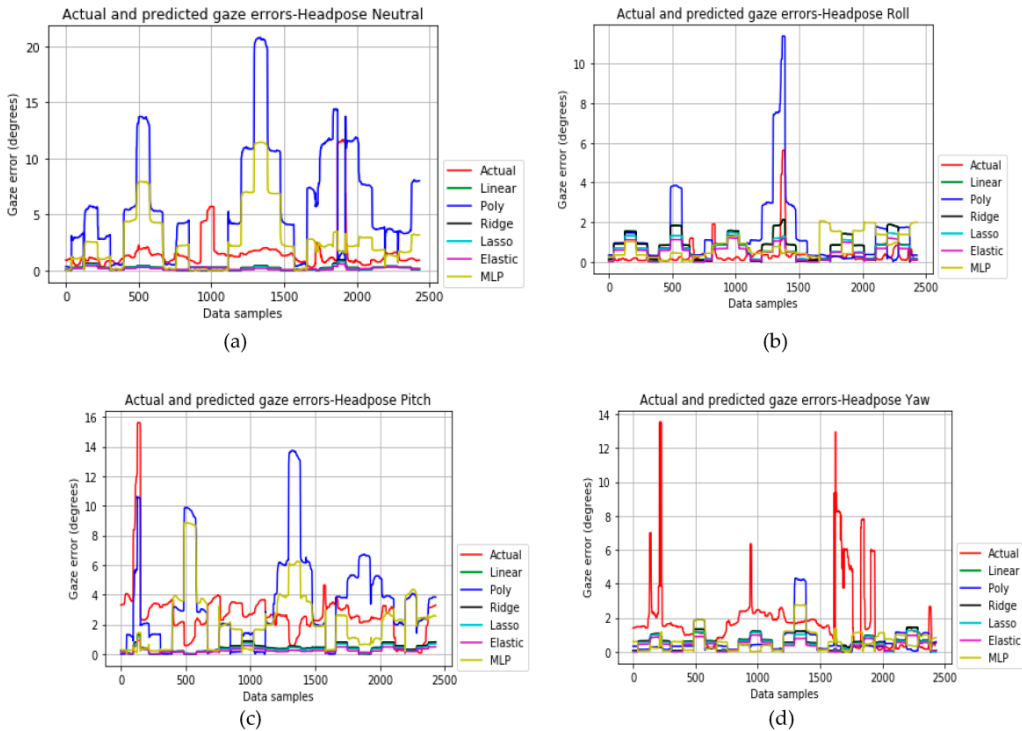

**Figure 15.** Actual (red) and predicted gaze errors for the head pose neutral, roll, pitch, and yaw (**a**–**d**) datasets.

Table 8 shows that ElasticNet (with penalty parameter = 0.5) has the lowest prediction errors for all head pose datasets. The regularization used for the Ridge and Lasso models is 0.001. The MLP model is used with 1 hidden layer with 100 units, ReLU activation, and a regularization value of 0.001.

*4.2. Platform Pose Error Model*

For the tablet pose dataset, the ElasticNet performs well (Figure 16a-d), with a penalty parameter value of 0.5. The MLP and polynomial regression models were seen to overestimate the error values. It was seen that outliers and noise strongly affect all the models and therefore outlier removal and data standardization methods to convert the input features to have a normal distribution were essential.

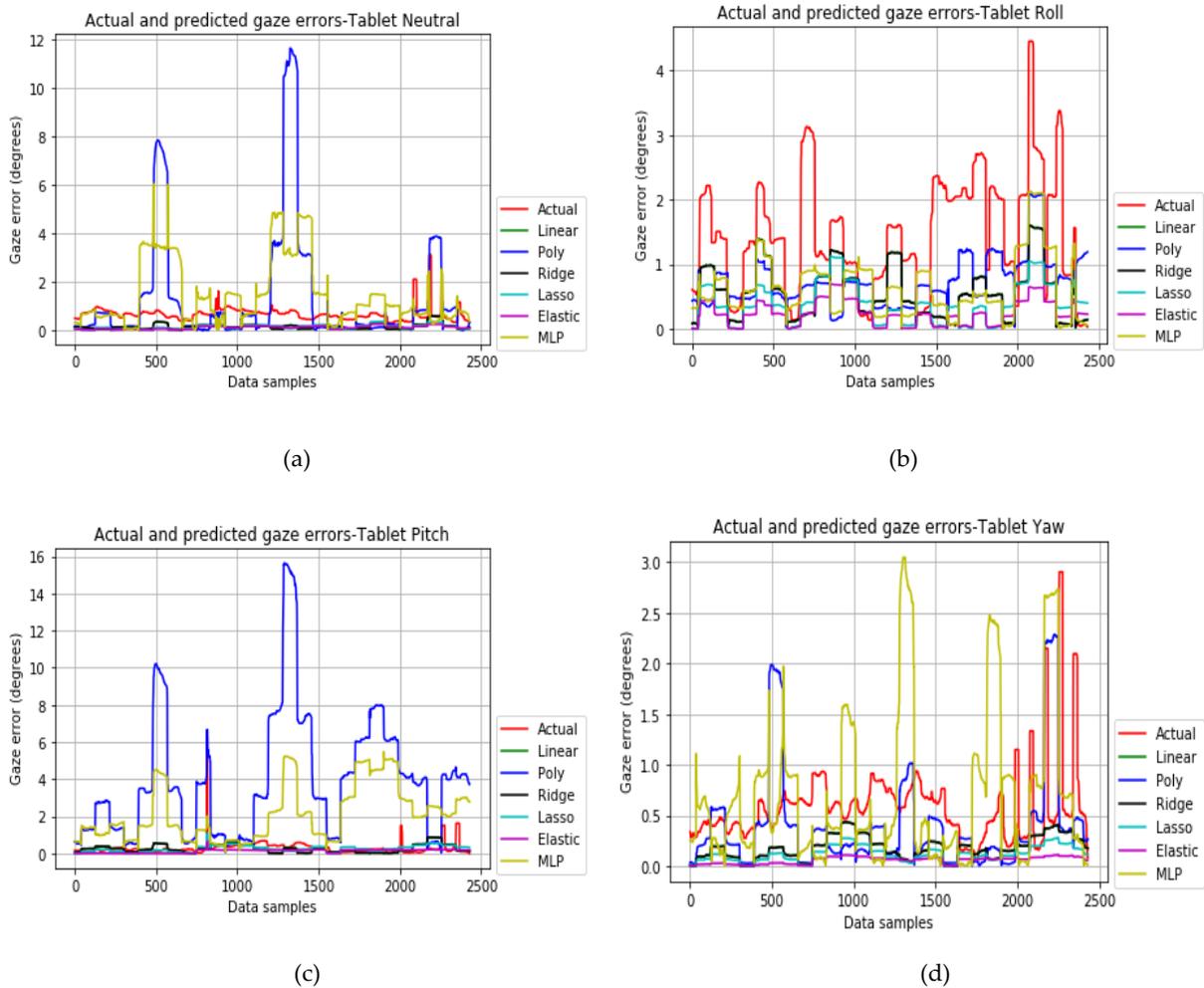

**Figure 16.** Actual gaze errors (red) and predicted gaze errors for the tablet pose neutral, roll, pitch, and yaw (**a**–**d**) datasets.

**Table 8.** Prediction errors (RMSE) for models of the desktop (left, red columns) and tablet (right, blue columns).

|                | Neutral | R20  | P20  | Y20  | Neutral | R20  | P20  | Y20  |
|----------------|---------|------|------|------|---------|------|------|------|
| Linear         | 2.24    | 1.36 | 1.45 | 2.71 | 0.73    | 1.79 | 0.58 | 0.72 |
| Polynomial     | 7.48    | 1.82 | 7.88 | 2.88 | 2.61    | 1.99 | 5.24 | 0.81 |
| Ridge          | 2.24    | 1.36 | 1.44 | 2.70 | 0.73    | 1.79 | 0.58 | 0.72 |
| Lasso          | 2.24    | 1.15 | 1.31 | 2.70 | 0.73    | 1.73 | 0.58 | 0.71 |
| ElasticNet     | 2.25    | 1.07 | 1.29 | 2.69 | 0.73    | 1.71 | 0.50 | 0.70 |
| Neural network | 4.01    | 1.21 | 2.09 | 2.75 | 1.75    | 1.78 | 2.39 | 1.18 |

*4.3. Establishment of the Error Models*

As described above, a regression model with three predictor variables can be expressed as:

$$Y = B_0 + B_1 * X_1 + B_2 * X_2 + B_3 * X_3, \qquad (29)$$

where $B_0$ is the intercept; $B_1$, $B_2$, and $B_3$ are coefficients; and $X_1$, $X_2$, and $X_3$ are input features (gaze, yaw, pitch angles). Since the ElasticNet model had the lowest RMSE for predicting the head and tablet pose errors, the $B_0$, $B_1$, $B_2$, and $B_3$ parameters of this model were computed for the head pose and tablet pose datasets and are presented in Table 9. With these parameters, the head and platform pose error models may be constructed for error prediction using gaze angular variables as the input.

**Table 9.** Coefficients and intercept of the best model for the head pose and platform pose.

| Platform/Condition | ElasticNet Coefficients $B_1$, $B_2$ and $B_3$ | Intercept $B_0$ |
|---|---|---|
| Desktop: Head pose neutral | [0.09336917, 0.19406989, −0.00279198] | −1.99912371e-16 |
| Desktop: Head Roll 20 | [0.0 , 0.65189252, 0.07303053] | 3.23942926e-16 |
| Desktop: Head Pitch 20 | [0.22606558, 0.11028886, 0.05731872] | −9.55229176e-17 |
| Desktop: Head Yaw 20 | [0.0 , 0.51352565, 0.08149052] | 8.94037155e-17 |
| Tablet: Platform pose neutral | [0.07333954, 0. , −0.17956056] | 1.76076146e-16 |
| Tablet: Platform Roll 20 | [0.0 , −0.31460996, −0.23620848] | −2.87637414e-16 |
| Tablet: Platform Pitch 20 | [0. 0, −0.05682588, −0.20804325] | 2.34007877e-16 |
| Tablet: Platform Yaw20 | [0.0 , −0.01596391, −0.06346607] | −2.41027682e-17 |

**5. MLGaze: An Open Machine Learning Repository for Gaze Error Pattern Recognition**

In this study, the major aim was to present the research direction on the pattern recognition of gaze errors and to show, as a proof of concept, that machine learning algorithms can be used to distinguish gaze error patterns created due to different operating conditions of an eye tracker. To test and replicate the principles, the full set of gaze data collected for this study was provided in the Mendeley data repository as mentioned in Section 2.5.5 above.

For any eye tracker other than the Eyetribe, the gaze error characteristics will be different based on the tracking principle and tracker hardware. Therefore, for other eye trackers, the classifier models described in this study have to be re-trained using data from the respective eye trackers if their characteristics vary widely from Eyetribe (which can be determined by comparing the respective eye tracker's data with the data provided in the Mendeley data repository). To facilitate this and make it possible for researchers to replicate the principles of this study on their own eye tracker's data, the codes for implementing the full data processing pipeline as described in this paper and in Figure 1 are provided in an open repository called MLGaze, hosted on GitHub. The link to the repository and its contents are discussed here: https://github.com/anuradhakar49/MLGaze.

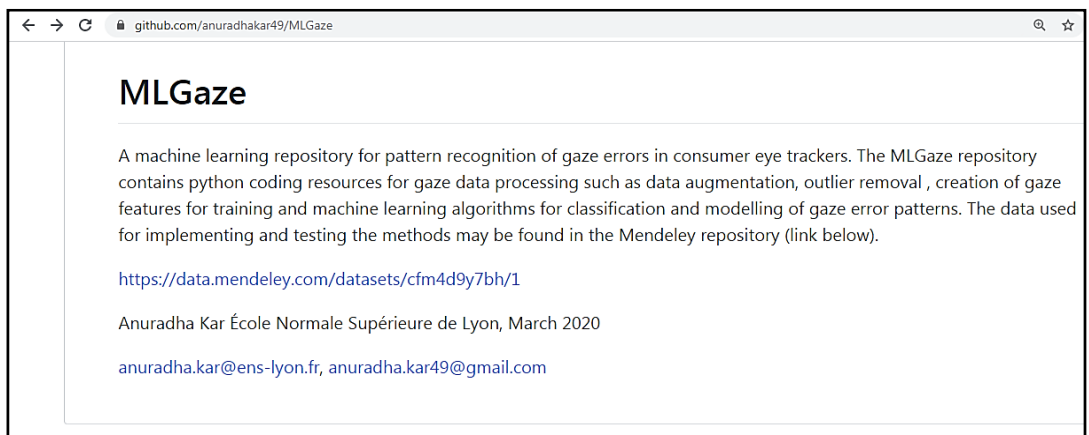

**Figure 17.** (**a**). The MLGaze repository hosted on GitHub, which contains Python coding resources for implementation of the methods described above. (**b**). Contents of the MLGaze repository.

There are eight python scripts in the repository, and these are fully documented. The codes in data_augmentation.py and outlier_removal.py may be used for data augmentation and outlier removal as described in Sections 3.2 and 2.4, respectively. Code create_features.py is for generating training features, whose output may be used with the codes train_SVM.py, train_KNN.py, and train_MLP.py to train KNN, SVM, and MLP models, respectively, as described in Sections 3.4 and 3.5. Code regression_models.py is for running all the regression models described in Section 4. The codes for ML models also produce the confusion matrices and detection scores as well as learning curves as described in Section 3 and Appendix B. Apart from this, sample CSV files are also included, which can be used to test the coding resources. The repository was released under the GNU GPLv3 license.

**6. Conclusions and Future Work**

In contemporary eye gaze research, there is a scarcity of analytical methods for studying the variability of gaze accuracy in eye tracking data acquired under unconstrained conditions. There is also a lack of datasets that provide gaze and ground truth information for different operating conditions, and gaze datasets labelled with different experimental scenarios are absent. Therefore, for this study, a new and diverse gaze dataset was collected and labelled with different operating conditions. The dataset named NUIG_EyeGaze01(Labelled eye gaze dataset) comprises of fixation data from 20 participants captured from different user platforms and operating conditions described in this paper. This dataset was then analyzed to identify the presence and to model the impacts of the above conditions on the gaze error levels.

Several new approaches with respect to gaze error pattern analysis were implemented in this study. This includes the use of multiple strategies for outlier removal, and it was found that the median filtering approach, among others, was successful in de-noising gaze data from all the collected datasets. Next, various gaze data augmentation methods were applied, which helped to increase the size of the collected gaze datasets by an order of magnitude and introduce a lot of variabilities within the datasets. The choice of gaze error features and use of a random forest-based feature selection method helped to reveal insights about the gaze error characteristics, e.g., that the interquartile range and confidence intervals are significant indicators for distinguishing gaze error patterns.

The t-SNE algorithm is a relatively new concept in machine learning, which was used in this study to visualize the distribution of classes within the gaze datasets, when gaze data influenced by different error sources were mixed together. This technique demonstrated the complexities of the gaze datasets used in this study, which are close to what might be expected in gaze data from unconstrained practical applications. Finally, the use of machine learning for classification and prediction of gaze error patterns in artificially created heterogeneous gaze datasets as done in this study is a new concept, which has been sparsely explored before.

The results from the training of the ML models reveal that these could be highly robust in identifying gaze error patterns even in complex gaze datasets. It was seen that classifier models can successfully distinguish between eye tracking data collected under normal conditions and data collected under varying operating conditions, such as a high degree of head pose variations, different user distances, and platform poses. With these classifiers, familiar error patterns in eye gaze datasets may be detected, and gaze error patterns that do not match the normal gaze behavior may be identified and recognized as new error types.

The various concepts developed in this study, including the application of classifier and regression models to gaze error data, may be used to build improved eye tracking systems and algorithms and also to obtain better results from them. The main benefit of training the classifier models is that they can distinguish anomalous gaze data present in realistic and complex gaze datasets as used in this study and possibly recognize what caused them (e.g., impact of user distance or head pose). This can help eye gaze application developers, researchers, or engineers to deploy appropriate prevention methods, compensation strategies, or setup improvements to reduce the impacts of the error sources affecting their data.

The regression models may be used predict how an eye tracker might behave under practical operating conditions like a high degree of head pose or platform pose variations and forecast the possible levels of error caused by different error sources. This can help eye gaze researchers or engineers to quantitatively specify the limits of their systems while operating under these challenging conditions and also develop suitable error correction methods.

Additionally, the outlier detection techniques described in this study could be useful for applications to any gaze dataset prior to analyzing them, to observe underlying data patterns. Overall, the gaze data analysis pipeline as implemented in this study could be of use to any eye gaze researcher or engineer who wishes to gain deeper insights into their collected gaze data and have an estimate about the impacts of various non-ideal operating conditions on their eye tracking system.

There are also several ways to extend the study presented here since extracting and analyzing error patterns from gaze data is a new and unexplored field in gaze research. For example, deep neural networks may be used for the detection of anomalous gaze data collected from dynamic eye tracking applications under unconstrained operating conditions. As mentioned in Section 1.2, in this study, the influence of one operating was considered at a time while collecting gaze data. However, complex scenarios may arise, where gaze data may be affected by more than one error source, such as, for example, in automotive use cases. The problem of a multi-factor influence on gaze data could be an interesting and complex research question but can only be answered when relevant and labelled data is available. Therefore, the collection of relevant gaze datasets from complex operational scenarios using desktop, handheld, head mounted, or automotive setups to understand a diverse range of gaze error patterns could be another potential future study in this domain.

**Supplementary Materials:** The eye gaze dataset collected from desktop and tablet platforms and used in this study is available online at: https://data.mendeley.com/datasets/cfm4d9y7bh/1 . A GitHub repository containing the Python implementations of the gaze error pattern recognition pipeline is made available online at: https://github.com/anuradhakar49/MLGaze .

**Funding:** The research work presented here was funded under the Strategic Partnership Program of Science Foundation Ireland (SFI) and co-funded by FotoNation Ltd. Project ID: 13/SPP/I2868 on "Next Generation Imaging for Smartphone and Embedded Platforms".

## Appendix A

The tablet was mounted on a gimbal holder as shown below in Figure A1(a). To measure the device angles, the data from the inertial measurement sensor of the tablet were used. These are highly accurate sensors found in most tablet/smartphones to estimate the orientation of the device with 0.5-1 degree of precision. There are apps like Accelerometer analyser for Windows tablets and smartphones, which can provide device orientation angles (in 3-D). Apart from this, the tablet orientation was verified by aligning a device (e.g., Android phone) with the tablet and running an Accelerometer app on the device to see if the two orientation results were similar. A screenshot from such an app (Tiltmeter, available on Google Play for Android) is shown in Figure A1(b). Mechanical protractors were occasionally used to manually verify the angles. Readings for one orientation were taken for all participants in consequent sessions to allow minimum tablet setup perturbation between the experiments.

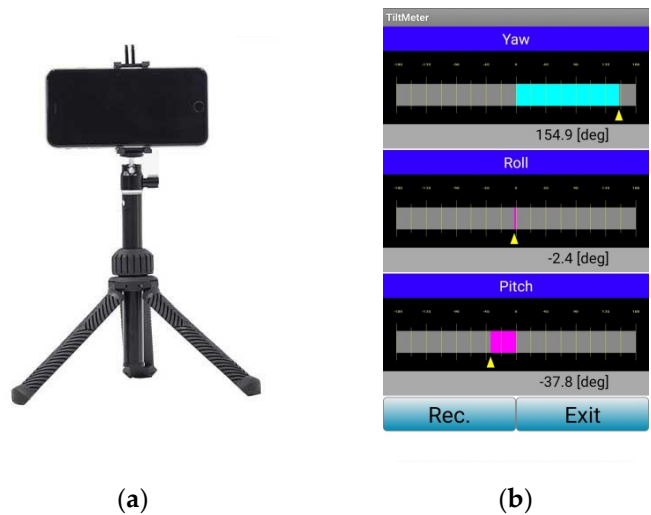

(**a**)   (**b**)

**Figure A1.** (**a**) Gimbal tripod mount for the tablet; (**b**) 3-D position reading from inertial sensors using an smartphone/tablet app.

## Appendix B

For this study, the selection of the participant number was based on the criteria about the number of data samples required compared to the size of the feature set used for training the machine learning (ML) algorithms. For the ML models like SVM, KNN, and MLP, there is an empirical rule that the training dataset size should be at least 10 times the size of the feature set. This is also termed as the "rule of 10" [63]. Going by this rule, it may be seen that in this paper, the maximum size of the feature set is 20 (Equation (15)) for the classification problem. Therefore, data from 20 persons with 10-fold augmentation were used. Additionally, the gaze yaw and pitch features were included in the training set, so the total training data size was (20 × 3 × 10) = 600 samples per class. As an example, for the classification of the user distance data from the desktop or tablet, there were four classes and all three categories of gaze data were collected for each of them. So, the total number of training samples was 600 × 4 = 2400 samples for the 4-class classification problem (this is shown in Table 3 above). This feature size was deemed to be useful while training and is supported by the learning curves for the different ML models whose plots are shown in Figure B1a–d below for the KNN, SVM, and MLP classifiers and the ElasticNet regression model, respectively. The training set size in the figures indicates the number of data samples used for training the respective model.

From the figures, it is seen that the training/cross validation accuracy for the different machine learning models converges and stabilizes into an asymptote within the training sample size for the classification and regression tasks and therefore it can be said that the training set size chosen is suitable for the machine learning model. Once the training and validation curves converge, it is unlikely that increasing the training data size will increase the model performance. While it is always possible to gather more data for studying the gaze error patterns, based on the trade-off between the available number of participants, the data collection time required, and the training/validation accuracy, the number of participants chosen in this study was found to be optimal.

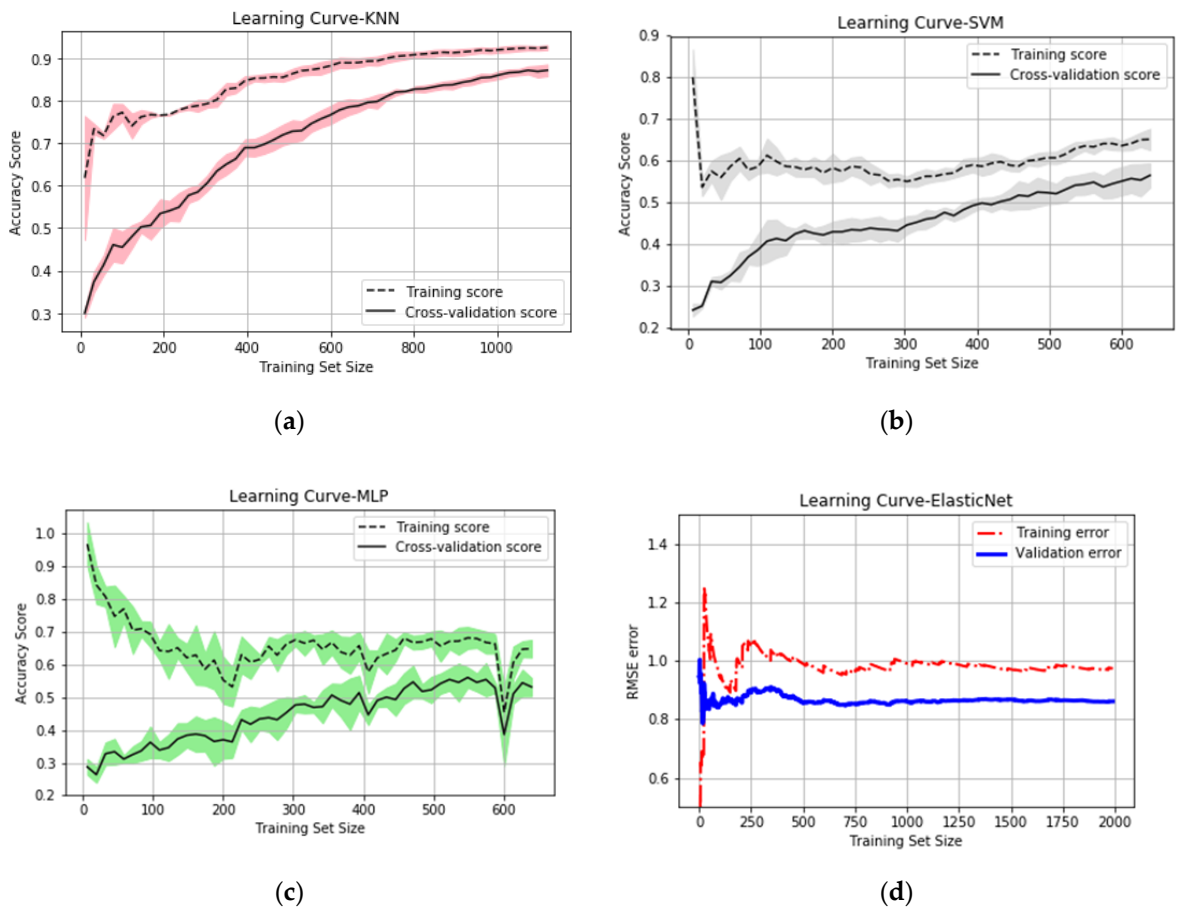

**Figure 1.** Learning curves for (**a**) KNN, (**b**) SVM, and (**c**) MLP classifiers, and the (**d**) Elastic-Net model.

**References**

1. Kar, A.; Corcoran, P. Performance Evaluation Strategies for Eye Gaze Estimation Systems with Quantitative Metrics and Visualizations. *Sensors* **2018**, *18*, 3151.

2. Cheon, M.; Lee, J. Gaze pattern analysis for video contents with different frame rates. In Proceedings of the *2013 Visual Communications and Image Processing (VCIP)*, Kuching, Malaysia, 17–20 November 2013; pp. 1–5.
3. Zhu, D.; Mendis, B.S.U.; Gedeon, T.; Asthana, A.; Goecke, R. A Hybrid Fuzzy Approach for Human Eye Gaze Pattern Recognition. In *Advances in Neuro-Information Processing*; ICONIP 2008; Lecture Notes in Computer Science; Köppen, M., Kasabov, N., Coghill, G., Eds.; Springer: Berlin/Heidelberg, Germany, 2009; Volume 5507.
4. Horiguchi, Y.; Suzuki, T.; Sawaragi, T.; Nakanishi, H.; Takimoto, T. Extraction and Investigation of Dominant Eye-Gaze Pattern in Train Driver's Visual Behavior Using Markov Cluster Algorithm. *2016 Joint 8th International Conference on Soft Computing and Intelligent Systems (SCIS) and 17th International Symposium on Advanced Intelligent Systems (ISIS), Sapporo*, Japan, August 25–28 2016; pp. 578–581.
5. Braunagel, C.; Geisler, D.; Stolzmann, W.; Rosenstiel, W.; Kasneci, E. On the necessity of adaptive eye movement classification in conditionally automated driving scenarios. In *Proceedings of the Ninth Biennial ACM Symposium on Eye Tracking Research & Applications* (ETRA 16); ACM: New York, NY, USA, 2016; pp. 19–26.
6. Koochaki, F.; Najafizadeh, L. Predicting Intention Through Eye Gaze Patterns. In *2018 IEEE Biomedical; Circuits and Systems Conference (BioCAS)*: Cleveland, OH, USA, 2018; pp. 1–4.
7. Pekkanen, J.; Lappi, O. A new and general approach to signal denoising and eye movement classification based on segmented linear regression. *Sci. Rep.* **2017**, *7*, 1–13.
8. Zemblys, R.; Niehorster, D.C.; Komogortsev, O.; Holmqvist, K. Using machine learning to detect events in eye-tracking data. *Behav. Res. Methods* **2018**, *50*, 160–181.
9. Barz, M.; Daiber, F.; Sonntag, D., and Bulling. A. Error-aware gaze-based interfaces for robust mobile gaze interaction. In *Proceedings of the 2018 ACM Symposium on Eye Tracking Research & Applications (ETRA '18)*; ACM: New York, NY, USA, 2018; Article 24, 10 p.
10. Barz, M.; Daiber, F.; Sonntag, D.; Bulling. A. Prediction of gaze estimation error for error-aware gaze-based interfaces. In *Proceedings of the Ninth Biennial ACM Symposium on Eye Tracking Research & Applications* (ETRA '16); ACM: New York, NY, USA, 2016; pp. 275–278.
11. Chandola, V.; Banerjee, A.; Kumar, V. Anomaly Detection: A Survey. *ACM Comput. Surv. (CSUR)* **2009**, *41*, 15:1–15:58.
12. Song, F.; Diao, Y.; Read, J.; Stiegler, A; Bifet, A. EXAD: A System for Explainable Anomaly Detection on Big Data Traces. *IEEE Intl. Conf. Data Mining Workshops*: Singapore, 2018; pp. 1435–1440.
13. Ishikawa, T.; Baker, S.; Matthews, I.; Kanade, T. Passive Driver Gaze Tracking with Active Appearance Models. *Proc. World Congr. Intell. Transp. Syst.* **2004**, 1–12, doi:10.1184/R1/6557315.v1.
14. Chen, T.; Ma, K.; Chen, L.H. Tri-state median filter for image denoising. In *IEEE Transactions on Image Processing*; IEEE: Piscataway, NJ, USA, 1999; Volume 8; no. 12; pp. 1834–1838.
15. Khalil, H.H.; Rahmat, R.O.K; Mahmoud, W.A. Chapter 15: Estimation of Noise in Gray-Scale and Colored Images Using Median Absolute Deviation (MAD). *In Proceedings of the 3rd International Conference on Geometric Modeling and Imaging*, London, UK, 9–11 July 2008; pp. 92–97.
16. Chen, Y.C. A tutorial on kernel density estimation and recent advances. *Biostat. Epidemiol.* **2017**, *1*, 161–187.
17. Koydemir, H.C.; Feng, S.; Liang, K.; Nadkarni, R.; Tseng, D.; Benien, P.; Ozcan, A. A survey of supervised machine learning models for mobile-phone based pathogen identification and classification. In *Proc. SPIE 10055, Optics and Bio-photonics in Low-Resource Settings III, 100550A (7 March 2017)*. International Society for Optics and Photonics: Washington, DC, USA, 2017.
18. Qiu, J.; Wu, Q.; Ding, G.; Xu, Y.; Feng, S. A survey of machine learning for big data processing. *EURASIP J. Adv. Sig. Process.* **2016**, *2016*, 67.
19. Bjerrum, E.J.; Glahder, M; Skov. T. Data Augmentation of Spectral Data for Convolutional Neural Network (CNN) Based Deep Chemometrics. *arXiv preprint* **2017**, arXiv:1710.01927.
20. Polson, N.G.; Scott, S.L. Data augmentation for support vector machines. *Bayesian Anal.* **2011**, *6*, 23, doi:10.1214/11-BA601.
21. Sáiz-Abajo, M.J.; Mevik, B.H.; Segtnan, V.H.; Næs, T. Ensemble methods and data augmentation by noise addition applied to the analysis of spectroscopic data. *Anal. Chim. Acta* **2005**, *533*, 147–159; ISSN 0003-2670.
22. Duchowski, A.; Jörg, S.; Allen, T.N.; Giannopoulos, I; Krejtz, K. Eye movement synthesis. In *Proceedings of the Ninth Biennial ACM Symposium on Eye Tracking Research & Applications* (ETRA '16); ACM: New York, NY, USA, 2016; pp. 147–154.

23. Duchowski, A.; Jörg, S.; Lawson, A.; Bolte, T.; Świrski, L; Krejtz, K. Eye movement synthesis with 1/f pink noise. In *Proc. 8th ACM SIGGRAPH Conf. on Motion in Games* (MIG '15); ACM: New York, USA, 2015; pp. 47–56.
24. Devries, T.; Taylor, G.W. Dataset Augmentation in Feature Space. *arXiv preprint* **2017**, arXiv:1702.05538.
25. Um, T.; Pfister, F.; Pichler, D.; Endo, S.; Lang, M.; Hirche, S.; Fietzek, U.; Kulić, D. Data augmentation of wearable sensor data for parkinson's disease monitoring using convolutional neural networks. In *Proceedings of the 19th ACM International Conference on Multimodal Interaction* (ICMI '17); ACM: New York, NY, USA, 2017; pp. 216–220.
26. Salamon, J.; Bello, J.P. Deep Convolutional Neural Networks and Data Augmentation for Environmental Sound Classification. *IEEE Signal Process. Lett.* **2017**, *24*, 279–283.
27. Li, Q.; Zhou, Y.; Chen, D. Research on machine learning algorithms and feature extraction for time series. *2017 IEEE 28th Annual International Symposium on Personal, Indoor, and Mobile Radio Communications (PIMRC)*, Montreal, QC, Canada, 8–13 October 2017; pp. 1–5.
28. Popescu, M.C; Sasu, L.M. Feature extraction, feature selection and machine learning for image classification: A case study. In Proceedings of the *2014 International Conference on Optimization of Electrical and Electronic Equipment (OPTIM)*, Bran, Romania, 22–24 May 2014; pp. 968–973.
29. Khalid, S.; Khalil, T.; Nasreen, S. A survey of feature selection and feature extraction techniques in machine learning. In Proceedings of the *2014 Science and Information Conference*, London, UK, 27–29 August 2014; pp. 372–378.
30. Oravec, M. Feature extraction and classification by machine learning methods for biometric recognition of face and iris. In Proceedings of the *ELMAR-2014*, , *Zadar*, Croatia, 10–12 September 2014; pp. 1–4.
31. van der Maaten, L.J.P.; Hinton, G.E. **Visualizing High-Dimensional Data Using t-SNE**. *J. Mach. Learn. Res.* **2008**, *9*, 2579–2605.
32. Rogovschi, N.; Kitazono, J.; Grozavu, N.; Omori, T; Ozawa, S. t-Distributed stochastic neighbor embedding spectral clustering. In Proceedings of the *2017 International Joint Conference on Neural Networks (IJCNN)*, Anchorage, AK, USA, 14–19 May *2017*; pp. 1628–1632.
33. Mounce, S. Visualizing Smart Water Meter Dataset Clustering With Parametric T-distributed Stochastic Neighbour Embedding. In Proceedings of the 2017 13th International Conference on Natural Computation, Fuzzy Systems and Knowledge Discovery (ICNC-FSKD), Guilin, China, 29–31 July *2017*; pp. 1940–1945.
34. Retsinas, G.; Stamatopoulos, N.; Louloudis, G.; Sfikas, G; Gatos, B. Nonlinear Manifold Embedding on Keyword Spotting Using t-SNE. In Proceedings of the *2017 14th IAPR International Conference on Document Analysis and Recognition (ICDAR)*, Kyoto, Japan, 9–15 November *2017*; pp. 487–492.
35. Pancerz, K.; Paja, W.; Gomuła, J. Random forest feature selection for data coming from evaluation sheets of subjects with ASDs. In Proceedings of the *2016 Federated Conference on Computer Science and Information Systems (FedCSIS)*, Gdansk, Poland, 11–14 September *2016*; pp. 299–302.
36. Cao, W.; Xu, J; Liu, Z. Speaker-independent speech emotion recognition based on random forest feature selection algorithm. In Proceedings of the *2017 36th Chinese Control Conference (CCC)*, Dalian, China, 26–28 July *2017*; pp. 10995–10998.
37. Gomes, R.; Ahsan, M.; Denton, A. Random Forest Classifier in SDN Framework for User-Based Indoor Localization. In Proceedings of the *2018 IEEE International Conference on Electro/Information Technology (EIT)*, Rochester, MI, USA, 3–5 May *2018*; pp. 0537–0542.
38. Song, Q.; Liu, X; Yang, L. The random forest classifier applied in droplet fingerprint recognition. In Proceedings of the *2015 12th International Conference on Fuzzy Systems and Knowledge Discovery (FSKD)*, Zhangjiajie, China, 15–17 August *2015*; pp. 722–726.
39. Okfalisa, I.; Gazalba, M; Reza, N.G.I. Comparative analysis of k-nearest neighbor and modified k-nearest neighbor algorithm for data classification. In Proceedings of the *2017 2nd International conferences on Information Technology, Information Systems and Electrical Engineering (ICITISEE)*, Yogyakarta, Indonesia, 1–2 November *2017*; pp. 294–298.
40. Guan, F.; Shi, J.; Ma, X.; Cui, W; Wu, J. A Method of False Alarm Recognition Based on k-Nearest Neighbor. In Proceedings of the *2017 International Conference on Dependable Systems and Their Applications (DSA)*, Beijing, China, 31 October–2 November *2017*; pp. 8–12.

41. Cai, Y.; Huang, H.; Cai, H; Qi, Y. A K-nearest neighbor locally search regression algorithm for short-term traffic flow forecasting. In Proceedings of the *2017 9th International Conference on Modelling, Identification and Control (ICMIC)*, Kunming, China, 10–12 July *2017*; pp. 624–629.
42. Zhang, X.; Li, B; Sun, X. A k-nearest neighbor text classification algorithm based on fuzzy integral. In Proceedings of the *2010 Sixth International Conference on Natural Computation*, Yantai, China, 10–12 August *2010*; pp. 2228–2231.
43. Waske, B; Benediktsson, J.A. Fusion of Support Vector Machines for Classification of Multisensor Data. In *IEEE Transactions on Geoscience and Remote Sensing*; *IEEE:* Piscataway, NJ, USA, 2007; Volume 45; no. 12, pp. 3858–3866.
44. Sheng, L.; Mengjun, W.; Lanyong, Z. Research on information fusion of infrared and radar sensor based on SVM. In Proceedings of the *2012 International Conference on Measurement, Information and Control*, Harbin, China, 18–20 May *2012*; pp. 98–101.
45. Nakano, T.; Nukala, B.T.; Zupancic, R., A.; Lie, D.Y.C.; Lopez, J.; Nguyen, T.Q. Gaits Classification of Normal vs Patients by Wireless Gait Sensor and Support Vector Machine (SVM) Classifier. *Int. J. Softw. Innov. (IJSI)* **2017**, *5*, 17–29.
46. Jeong, G.; Truong, P.H.; Choi, S. Classification of Three Types of Walking Activities Regarding. *IEEE Sens. J.* **2017**, *17*, 2638–2639,.
47. Bengio, Y.; Courville,A.; Vincent, P. Representation Learning : A Review and New Perspectives. *IEEE Trans. Pattern Anal. Mach. Intell.* **2013**, *35*, 1798–1828.
48. Koldowski, U.M.M. Spiking neural network vs multilayer perceptron : Who is the winner in the racing car computer game. *Soft Comput.* **2015**, *12*, 3465–3478.
49. Bieniasz, J.; Rawski, M.; Skowron, K.; Trzepiński, M. Evaluation of multilayer perceptron algorithms for an analysis of network flow data. In *Photonics Applications in Astronomy, Communications, Industry, and High-Energy Physics Experiments 2016*; International Society for Optics and Photonics: Washington, DC, USA, 2016.
50. Lang, B. Monotonic Multi-layer Perceptron Networks as Universal Approximators. In *Artificial Neural Networks: Formal Models and Their Applications–ICANN 2005*; ICANN 2005; Lecture Notes in Computer Science; Duch, W., Kacprzyk, J., Oja, E., Zadrożny S., Eds.; Springer: Berlin/Heidelberg, Germany, 2005; Volume 3697.
51. Astorino, A.; Fuduli, A. The Proximal Trajectory Algorithm in SVM Cross Validation. *IEEE Trans. Neural Networks Learn. Syst.* **2016**, *27*, 966–977.
52. Huang, Q.; Mao, J; Liu, Y. An Improved Grid Search Algorithm of SVR Parameters Optimization. In Proceedings of the 2012 IEEE 14th International Conference on Communication Technology, Chengdu, China, 9–11 November 2012; pp. 1022–1026.
53. Chen, S.; Liu, C. Eye detection using discriminatory Haar features and a new efficient SVM. *Image Vis. Comput.* **2015**, *33*, 68–77.
54. Shang, W.; Cui, J.; Song, C.; Zhao, J; Zeng, P. Research on Industrial Control Anomaly Detection Based on FCM and SVM. In Proceedings of the *12th IEEE International Conference* on *Big Data Science* and *Engineering*, TrustCom/BigDataSE, New York, NY, USA, 1–3 August 2018; pp. 218–222.
55. Xie, Y.; Zhang, Y. An intelligent anomaly analysis for intrusion detection based on SVM. In Proceedings of the *2012 International Conference on Computer Science and Information Processing (CSIP)*, Xi'an, China, 24–26 August *2012*; pp. 739–742.
56. Guang, Y.; Min, N.I.E. Anomaly Intrusion Detection Based on Wavelet Kernel LS-SVM. In Proceedings of the 2013 3rd International Conference on Computer Science and Network Technology, Dalian, China, 12–13 October 2013; pp. 434–437.
57. Eswaran, C.; Logeswaran, R.A. Comparison of ARIMA, Neural Network and Linear Regression Models for the Prediction of Infant Mortality Rate. In *2010 Fourth Asia International Conference on Mathematical/Analytical Modelling and Computer Simulation*; IEEE: Piscataway, NJ, USA, 2010; pp. 34–39.
58. Friedman, J.; Hastie, T.; Tibshirani, R. Regularization Paths for Generalized Linear Models via Coordinate Descent. *J. Stat. Softw.* **2010**, *33*, 1–22.
59. Jones, L.K. Local Minimax Learning of Functions with Best Finite Sample Estimation Error Bounds: Applications to Ridge and Lasso Regression, Boosting, Tree Learning, Kernel Machines, and Inverse Problems. *IEEE Trans. Inf. Theory* **2009**, *55*, 5700–5727.

60. Kirpich, A.; Ainsworth, E.A.; Wedow, J.M.; Newman, J.R. B.; Michailidis, G.; McIntyre, L.M Variable selection in omics data: A practical evaluation of small sample sizes. *PLoS ONE* 2018, 13, e0197910.
61. Bayindir, R.; Gok, M.; Kabalci, E; Kaplan, O. An Intelligent Power Factor Correction Approach Based on Linear Regression and Ridge Regression Methods. In Proceedings of the 2011 10th International Conference on Machine Learning and Applications and Workshops, Honolulu, Hawaii, USA, 18–21 December 2011; pp. 313–315.
62. Verma, T.; Tiwana, A.P.S.; Reddy, C.C. Data Analysis to Generate Models Based on Neural Network and Regression for Solar Power Generation Forecasting. In Proceedings of the 2016 7th international conference on intelligent systems, modelling and simulation (ISMS), Bangkok, Thailandpp, 25–27 January 2016; pp. 97–100.
63. Vabalas, A.; Gowen, E.; Poliakoff, E.; Casson, A. () Machine learning algorithm validation with a limited sample size. *PLoS ONE* 2019, 14, e0224365.